\documentclass[journal]{IEEEtran}
\IEEEoverridecommandlockouts

\usepackage{cite}
\usepackage{multicol}

\usepackage{amssymb}

\usepackage{amsmath}
\usepackage{amsfonts}
\usepackage[T1]{fontenc}
\usepackage{algpseudocode}
\usepackage[utf8]{inputenc}
\usepackage[english]{babel}
\usepackage[usenames, dvipsnames]{color}
\usepackage{pifont}
\usepackage{booktabs}
\usepackage{mathtools}
\usepackage{textcomp}
 \usepackage{graphicx}
\usepackage{subfigure}
\usepackage[ruled,vlined,commentsnumbered]{algorithm2e}
\usepackage{fixltx2e}
\usepackage{psfrag}
\usepackage{grffile}
\usepackage{amsthm}
\usepackage{setspace}
\usepackage{diagbox}
\usepackage{graphicx} 
\usepackage{multirow}

\SetKwInput{KwIn}{Inputs}

\theoremstyle{definition}

\theoremstyle{remark}

\theoremstyle{plain}

\newcommand{\RNum}[1]{\uppercase\expandafter{\romannumeral #1\relax}}

\usepackage{setspace}

\def\BibTeX{{\rm B\kern-.05em{\sc i\kern-.025em b}\kern-.08em
    T\kern-.1667em\lower.7ex\hbox{E}\kern-.125em}}

\begin{document}

\title{When Next-Gen Sensing Meets Legacy Wi-Fi: Performance Analyses of IEEE 802.11bf and 
\\IEEE 802.11ax Coexistence}
\author{Navid~Keshtiarast,~\IEEEmembership{Graduate~Student~Member,~IEEE,}
        ~Pradyumna~Kumar~Bishoyi,~\IEEEmembership{Member,~IEEE,}
        ~Ido Manuel Lumbantobing,
	and~Marina~Petrova,~\IEEEmembership{Member,~IEEE}
    \thanks{Part of this work was accepted for presentation at the 2025 IEEE International Conference on Communications (ICC).}
    \thanks{This work was supported by the German Federal Ministry of Education and Research (BMBF) within the project “Open6GHub” under grant 16KISK012.
   
   All the authors are with Mobile Communications and Computing Group, at RWTH Aachen University, Germany (e-mail: \{navid.keshtiarast@mcc, pradyumna.bishoyi@mcc, ido.lumbantobing, petrova@mcc\}.rwth-aachen.de).}

}
\maketitle

\begin{abstract}

Sensing is emerging as a vital future service in next-generation wireless networks, enabling applications such as object localization and activity recognition. The IEEE 802.11bf standard extends \mbox{Wi-Fi} capabilities to incorporate these sensing functionalities. 
However, coexistence with legacy \mbox{Wi-Fi} in densely populated networks poses challenges, as contention for channels can impair both sensing and communication quality. This paper develops an analytical framework and a system-level simulation in \mbox{ns-3} to evaluate the coexistence of IEEE 802.11bf and legacy 802.11ax in terms of sensing delay and communication throughput. For this purpose, we have developed a dedicated ns-3 module for IEEE 802.11bf, which is made publicly available as open-source. We provide the first coexistence analysis between IEEE 802.11bf and IEEE 802.11ax, supported by link-level simulation in \mbox{ns-3} to assess the impact on sensing delay and network performance. Key parameters, including sensing intervals, access categories, network densities, and antenna configurations, are systematically analyzed to understand their influence on the sensing delay and aggregated network throughput. The evaluation is further extended to a realistic indoor office environment modeled after the 3GPP TR 38.901 standard. Our findings reveal key trade-offs between sensing intervals and throughput and the need for balanced sensing parameters to ensure effective coexistence in \mbox{Wi-Fi} networks.

\end{abstract}

\section{Introduction} \label{sec:Introduction}
Integrated sensing and communication (ISAC) will be a major technology enabler for new services in the next-generation wireless networks \cite{liu2020,nokia2023,Survey_acc}. \mbox{Wi-Fi} networks, due to their widespread popularity and ultra-dense deployment, are an ideal platform for large-scale ISAC applications \cite{TGbf_nist,Du_comst}. In September 2020 the IEEE 802.11 \mbox{Wi-Fi} working group established a new Task Group, namely IEEE 802.11bf (TGbf), to specify the essential modification to the existing physical and medium access control (PHY and MAC) layers to support sensing \cite{bf-draft,Restuccia_mag}. IEEE 802.11bf will operate in both sub-7 (2.4 GHz, 5 GHz, and 6 GHz) and mmWave (above 45 GHz) unlicensed frequency bands and is planned to support diverse applications ranging from passive localization of objects/humans to activity recognition by capturing micro-doppler signature \cite{nist_iotj,widmer_tmc} and up to 3D mapping of indoor environment \cite{nist_icassp}. 

The PHY of IEEE 802.11bf reuses the existing features of IEEE 802.11 ac/ax for sub-7 GHz and IEEE 802.11 ad/ay for mmWave and introduces additional processing modules for sensing operation \cite{widmer_2020,nist_mmwave}. In terms of the MAC layer, it is expected to adhere to the listen-before-talk (LBT) protocol for channel access, following the other wireless technologies operating in the unlicensed bands such as 5G new radio unlicensed (NR-U) \cite{navid_lnet2024}, ZigBee, and Bluetooth. However, certain modifications are necessary to enhance the sensing framework and ensure co-existence with the widely deployed legacy \mbox{Wi-Fi} networks in the sub-7 GHz bands, which is the primary focus of our work. Consequently, the design and analysis of an appropriate MAC layer protocol for IEEE 802.11bf networks is an emerging topic of high importance for the success of \mbox{Wi-Fi} sensing in sub-7 GHz unlicensed bands.

\subsection{Motivation}
 In an 802.11bf network, when an access point (AP) intends to perform sensing, it has to adopt carrier sense multiple access with collision avoidance (CSMA/CA) protocols to contend for the channel with other 802.11bf APs and legacy \mbox{Wi-Fi} stations \cite{peter_icassp,TGbf_nist}.
 As the number of contending nodes increases, the probability of accessing the channel for sensing decreases. This intermittent access to the channel results in irregular acquisition of sensing measurements and coarse-grained sensing accuracy \cite{peter_jstsp}. Further, in applications involving dynamic environment, e.g., moving target detection, this sporadic access to the channel significantly increases the position estimation error. In contrast, frequent channel acquisition for sensing, adversely affects the communication network performance. Balancing these conflicting objectives demands a thorough understanding of how MAC-layer mechanisms and key network parameters influence overall coexistence outcomes.

Despite initial studies on 802.11bf, there remains a gap in quantifying how its sensing MAC protocol affects the performance of 802.11ax, and current simulation tools do not fully capture the nuances of \mbox{Wi-Fi} sensing MAC protocols.
Furthermore, no scenario-driven examination currently explores how varying network configurations, such as AP density, channel bandwidth, antenna array size, and sensing intervals, can each introduce unique performance bottlenecks. Additionally, prioritizing the 802.11bf sensing traffic by varying access categories can improve the sensing performance at the expense of legacy communication throughput. Our work tackles these issues by presenting an analytical model that captures how coexistence impacts both sensing latency and legacy throughput, coupled with a novel ns-3 simulation module that implements 802.11bf on top of the standard 802.11ax framework.

\subsection{Related Works}
In recent years, very few works\cite{grossi_tsp,munari_pimrc,navid_wcnc2024} have addressed the coexistence of sensing and communication networks in unlicensed bands by studying the effect of mutual interference from a PHY layer perspective, using stochastic geometry approach. While stochastic geometry provides valuable insights into large-scale interference patterns and helps quantify coverage or outage probabilities, it generally does not account for MAC-layer functionalities, such as distributed coordination and channel access strategies that significantly influence coexistence performances.

A recent work in \cite{sahoo2024}, has analyzed the sensing performance of an 802.11bf network from MAC layer perspective and discussed the impact of different channel contention protocols to obtain the initial channel access for sensing.
In particular, the authors consider AP as a sensing initiator under two channel access methods, namely enhanced distributed channel access (EDCA) and point coordination function (PCF) interframe space (PIFS). The performance of both channel access methods is compared in terms of sensing overhead and the percentage of missed sensing instances for different sensing application requirements, and the simulation results reveal that PIFS outperforms EDCA since the PIFS gives priority access to the channel. We note that this study confirming the advantage of PIFS-based access, has been performed in a single 802.11bf network scenario without consideration of legacy Wi-Fi communication networks. In the case of a coexistence scenario with 802.11ax networks, prioritizing access for the 802.11bf network over the 802.11ax network will result in a significant unfair advantage in terms of channel access and will be more disruptive to 802.11ax users.

\subsection{Contributions}
In this paper, we bridge the above gap and investigate the performance of 802.11bf networks as well as their coexistence with 802.11ax communication networks in terms of communication throughout and sensing delay. The main contributions of our work are:

\begin{itemize}
    \item First, we analyze and determine \textit{the sensing delay that 802.11bf network is going to endure in case of coexistence, and the cost of coexistence in terms of throughput to the legacy \mbox{Wi-Fi} network}. We develop an analytical framework that quantifies the performance of both networks. 
    \item Second, we design and implement a \textit{system-level simulation framework using network simulator 3 (\mbox{ns-3}) for 802.11bf network}. The code is implemented on \mbox{ns-3.40} and includes the following prominent features to integrate the sensing functionality in the existing 802.11ax module: (i) point coordination function (PCF) on top of 802.11ax's EDCA for channel access and (ii) Channel sounding to perform sensing measurement instances. To the best of our knowledge, this is the first network simulator framework for analyzing 802.11bf networks. We made our framework code available\cite{bf_ns3} to the research community to enable reproducibility and support further research on the performance evaluation of 802.11bf in sub-7 GHz. 
    \item Finally, we validate the 802.11bf \mbox{ns-3} module output with the analytical model. We consider two different environment layouts: the first is a generic layout, and the second is an indoor office environment (as specified in 3GPP TR 38.901\cite{3GPP2}). Through extensive simulations, we analyze coexistence performance in terms of sensing latency and throughput, as well as the tradeoff between them by tuning various MAC parameters. These parameters include assigning different access categories to 802.11bf, adjusting sensing intervals, varying available channel bandwidth and altering the number of 802.11bf and 802.11ax APs.
\end{itemize}

The remainder of this paper is organized as follows: Section II provides a primer on the 802.11bf standard. In Section III, we present our analytical model for coexistence performance analyses. Section IV presents the 802.11bf implementation on \mbox{ns-3}. Section V offers a comprehensive analysis of the simulation results and discusses their implications. Finally, Section VI concludes the paper.
\section{Primer on 802.11bf}\label{sec:primer}

\begin{figure}
  \centering
  \includegraphics[width=0.44\textwidth,trim = 5mm 18mm 5mm 12mm,clip]{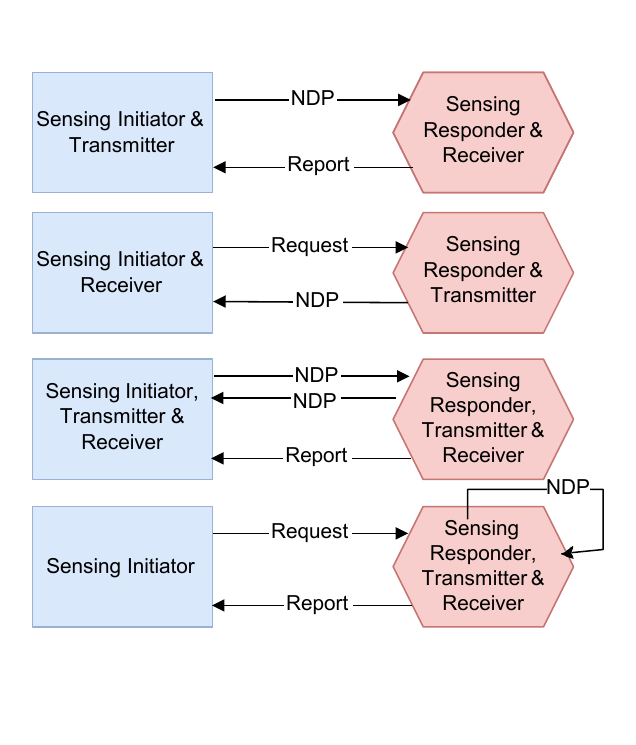}
  \caption{Illustration of roles and sensing configuration of 802.11bf.}\label{fig:roles_and_configuration}
\end{figure}
The primary objective of the 802.11bf standard is to establish a unified and standardized WLAN sensing operation for acquisition of sensing measurements. For that, the standard defines two roles for APs and stations (STAs). The first one is sensing  \textit{initiator and responder}, wherein the participant that initiates the sensing, either an AP or STA, is termed the initiator, while the participant that responds is designated as the responder. The second role is the sensing \textit{transmitter and receiver}, where the participant transmitting the null data packet (NDP) is the transmitter, and the participant receiving the NDPs and performing sensing measurements is termed the receiver. Figure \ref{fig:802.11ax_bf_coexist} shows all possible role combinations. A sensing initiator can be either a sensing transmitter or receiver, both or neither. For instance, in the first scenario (in Figure \ref{fig:802.11ax_bf_coexist}), the sensing initiator functions as a transmitter, transmitting the NDPs, while the sensing responder reports back the downlink channel state information (CSI) measurement. In the second scenario, the initiator functions as a receiver, requests the sensing responder to transmit the NDPs, and performs the uplink CSI measurement itself. Unlike the previous two scenarios, in the third case, the sensing initiator functions as both transmitter and receiver. This is mostly used to gather both uplink and downlink CSI measurements. In the fourth scenario, the sensing initiator is neither a transmitter nor a receiver. In this case, the sensing responder reports back the stored sensing information to the initiator, referred to as the sensing by proxy (SBP) procedure in the 802.11bf standard \cite{bf-draft}. Analogous to the functions of the sensing initiator, the sensing responder can be either the sensing transmitter or receiver or both during a sensing measurement procedure.

The IEEE 802.11bf sensing procedure in sub-7 GHz, which is the main focus of this work, distinguishes two broad categories based on the use of trigger frames to initiate the sensing measurement. One is \textit{trigger-based (TB) configuration}, where the sensing initiator is the AP and the responders are one or multiple STAs. The other one is \textit{non-TB configuration}, where the STA is the sensing initiator and the AP is the responder. In this work, we focus on TB configuration. In the following subsections, we first discuss how the AP obtains access to the channel for performing sensing measurement; then we describe the phases in the TB configuration-based sensing measurement instance.

\begin{figure}[t]
\centering\includegraphics[width=0.47\textwidth,trim = 0mm 0mm 0mm 0mm,clip]{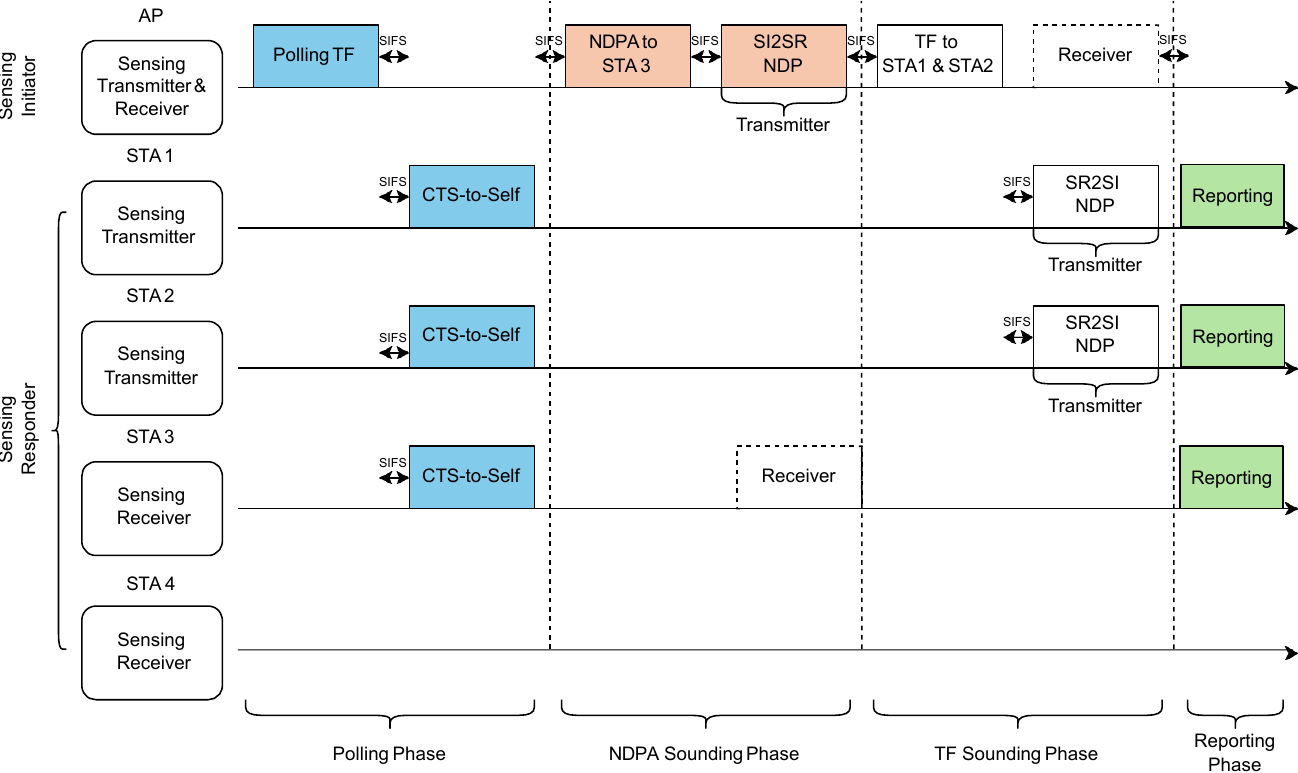}
	\caption{Trigger-based sensing measurement protocol.}
	\label{fig:System_model1}
\end{figure}

\subsection{Initial Channel Access}
Prior to initiating the TB-based measurement, the sensing transmitter AP first adheres to a legacy CSMA/CA protocol to access the channel. In particular, the AP as sensing initiator contends for the transmission opportunity (TXOP) with other contending sensing and communication APs using the EDCA mechanism as introduced in IEEE 802.11e standard \cite{802.11e}. The EDCA differentiates services using different priorities and maps them to access category (AC) with AC-specific contention window (CW) sizes, arbitrary interframe space (AIFS) values, and TXOPs. The EDCA parameters of different ACs are provided in Table \ref{tab:edca_parameters}. The higher-priority-ACs are assigned lower AIFS and smaller CW ranges to facilitate prioritized access to the channel. The effect of different ACs on the overall sensing performance in shown in Section V-C.

\subsection{Phases in TB Configuration}\label{subsec:TB_config}
TB configuration involves four phases, namely polling phase, null data packet (NDP) Announcement (NDPA) sounding phase, Trigger Frame (TF) sounding phase, and Reporting phase, as described below (see also Figure \ref{fig:System_model1}).

\begin{itemize}
    \item \textit{Polling Phase}: The AP starts the procedure by broadcasting the polling trigger frame (TF). The sensing receivers, in this case STAs, respond if available by transmitting clear to send (CTS)-to-self frame following a short inter-frame space (SIFS) period. Using OFDMA, multiple STAs can transmit their CTS-to-self frame simultaneously. The STA can refrain from participating in the sensing process by not responding to the polling TF frame (e.g., STA 4 in Figure \ref{fig:System_model1}). 
    \item \textit{NDPA sounding phase}: In this phase, the AP transmits the NDPA frame to the intended receiver to initiate the measurement. Thereafter, following an interval of SIFS, the AP transmits the sensing initiator to the sensing responder (SI2SR) NDP to perform downlink-based sensing.
    \item \textit{TF sounding phase}: The roles of transmitter and receiver are then reversed and the sensing is done in uplink transmission in this phase. Mutilple STAs transmit the responder to sensing initiator (SR2SI) NDP frame to the AP. To accommodate multiple STAs, the AP transmits a TF frame to allocate resources and timing prior to the transmission of SR2SI frame.
    \item \textit{Reporting phase}: The STAs transmit the sensing measurement results in a reporting feedback frame to the AP. The reporting frame mainly consists of CSI collected by the STAs during the NDPA sounding phase. The AP may use OFDMA or multi-user multiple input multiple output (MU-MIMO) to schedule the simultaneous uplink of multiple STAs.
\end{itemize}
The sensing procedure of IEEE 802.11bf protocol starts with the AP's initial channel access process, followed by the four phases of the sensing measurement session. The AP's probability of getting access to the channel depends on its EDCA parameters and the number of contending APs. Moreover, in a coexistence scenario, the number of APs contending for the channel would increase, leading to a higher initial channel access delay. In the following section, we develop an analytical framework and detailed performance analysis of sensing delay in the 802.11bf network and aggregated throughput in the 802.11ax network.

\begin{table}[htbp]
    \centering
    \caption{EDCA Parameters for IEEE 802.11 Access Categories}
    \label{tab:edca_parameters}
    \begin{tabular}{@{}lccc@{}}
        \toprule
        \textbf{Access Category} & \textbf{CW$_{\text{min}}$} & \textbf{CW$_{\text{max}}$} & \textbf{AIFS} \\ \midrule
        Background (BK)          & 15                        & 1023                       & 7            \\
        Best-Effort (BE)         & 15                        & 1023                       & 3            \\
        Video (VI)               & 7                         & 15                         & 2            \\
        Voice (VO)               & 3                         & 7                          & 2            \\ \bottomrule
    \end{tabular}
\end{table}

 \begin{figure}[t]
\centering\includegraphics[width=0.49\textwidth,trim = 45mm 15mm 40mm 4mm,clip]{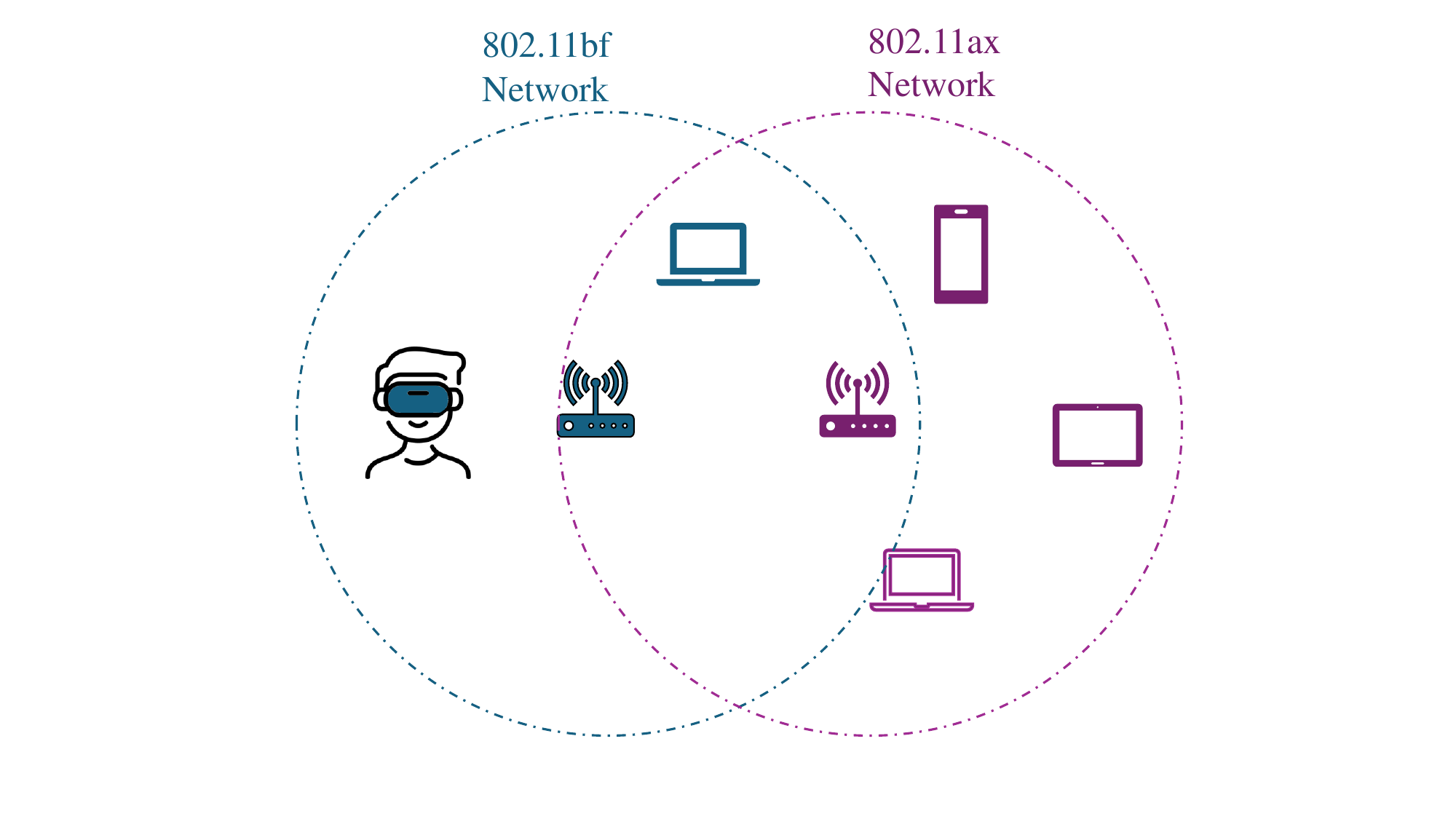}
	\caption{Coexistence of 802.11bf and 802.11ax networks.}
	\label{fig:802.11ax_bf_coexist}
\end{figure}

\section{Analytical Model for Coexistence Analysis}\label{sec:SystemModel}
We consider $N_{bf}$ 802.11bf and $N_{ax}$ 802.11ax APs sharing the same unlicensed spectrum, as shown in the Figure~\ref{fig:802.11ax_bf_coexist}. There are $N$ sensing STAs connected to each 802.11bf AP and  $M$ STA associated with each 802.11ax AP. We assume that all the APs and STAs are exposed to one another, i.e. are located within the range where energy detection is possible. The CSMA/CA scheme with binary slotted exponential backoff is adopted for both the 802.11bf and 802.11ax networks. We consider aggregated throughput as the Key Performance Indicator (KPI) for the communication network, while the average sensing delay is the KPI for the 802.11bf network. We assumed that the 802.11ax APs have saturated traffic, while the 802.11bf APs continuously request to perform sensing procedures. Further, we extend our sensing delay analysis to account for in-frequent sensing request scenario, where the 802.11bf AP perform sensing with various sensing repetition intervals, as specified in the 802.11bf standard\cite{bf-draft, sahoo2024}.

\subsection{Channel Access Mechanism}
\begin{figure}[h!]
\centering\includegraphics[width=0.49\textwidth,trim = 4mm 2mm 20mm 0mm,clip]{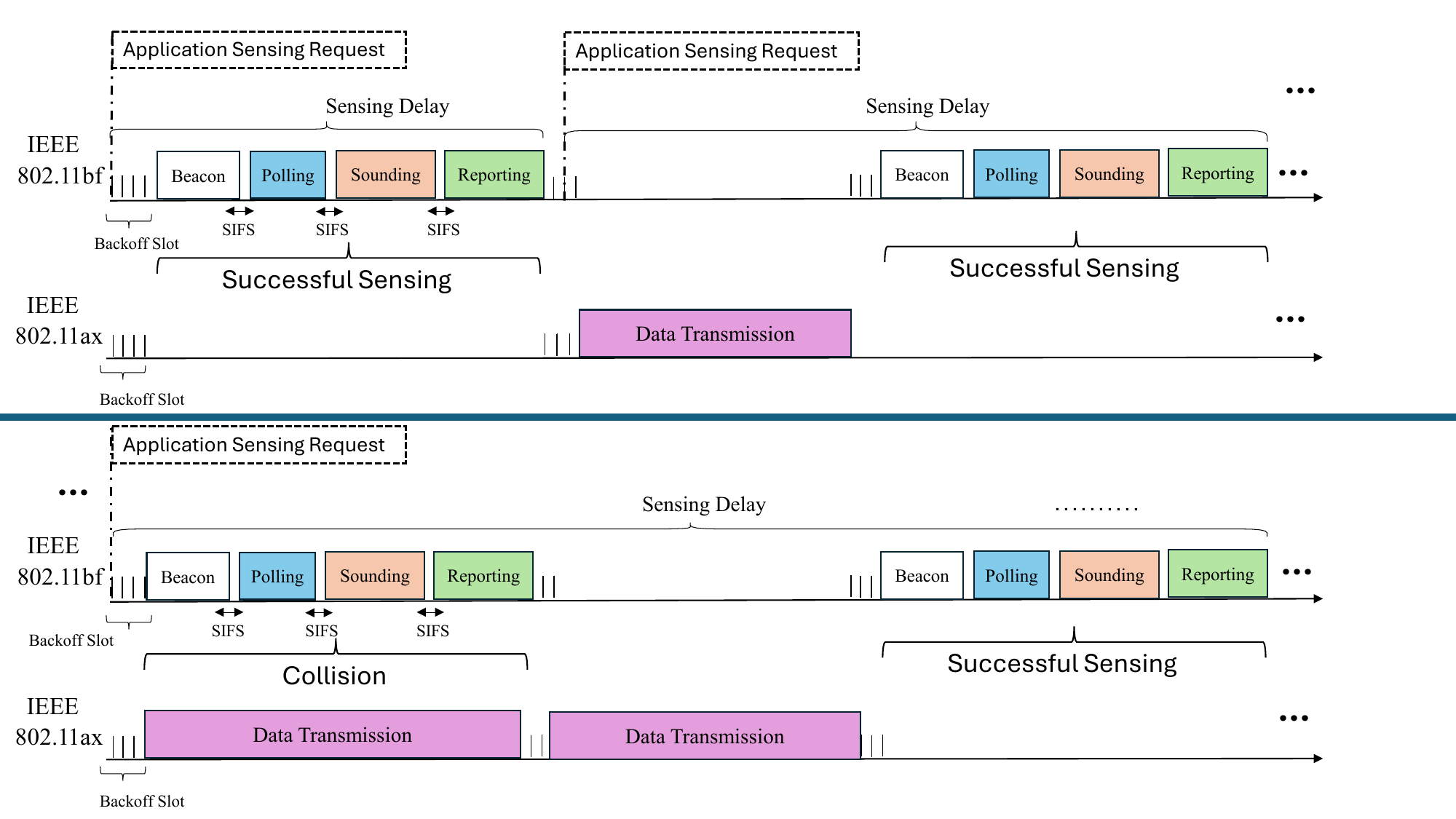}
	\caption{Illustration of 802.11bf and 802.11ax AP operation in a coexistence scenario.}
	\label{fig:System_model1_coexist}
\end{figure}

Figure~\ref{fig:System_model1_coexist} illustrates the channel access mechanism of both 802.11ax and 802.11bf networks in a coexistence scenario. Initially, both technologies contend for channel access using the EDCA mechanism as discussed in Section~\ref{sec:primer}. When an 802.11ax AP successfully accesses the channel, it can hold the channel for a Transmission Opportunity (TXOP) duration, which, according to the 802.11ax standard \cite{802.11ax_Khorov}, can be up to 5.484~ms when using aggregated A-MPDU frame transmissions. Conversely, if an 802.11bf AP gains access, it initiates the sensing procedure. 
In Figure~\ref{fig:System_model1_coexist}, we present two scenarios: one without collisions, allowing uninterrupted channel access, and another with collisions. In the collision scenario, retransmissions are required, which lead to increased sensing delays.

\subsection{Derivation of Probability of Collision and Success}
We assume that the time is divided into MAC slots. Since both 802.11ax and 802.11bf APs MAC employ CSMA/CA, we describe the behaviour of each node using the well-established Bianchi's two-dimensional discrete-time Markov chain \cite{840210, Ekici} and modify it to include the effect of coexistence of two heterogeneous networks.

Let $\tau_{ax}$, $\tau_{bf}$ define the attempt probability of an 802.11ax AP and 802.11bf in a given slot, respectively. The expression for $\tau_{bf}$ is given as
\begin{align}\label{coll1_prob}
    \tau_{bf}=\frac{(1-P_{bf}^{L+1})/(1-P_{bf})}{\left(\sum_{j=0}^{L}\left[1+\frac{1}{1-P_f} \sum_{k=1}^{W_j^{bf}-1} \frac{W_j^{bf}-k}{W_j^{bf}}\right] P_{bf}^j\right)},
\end{align}
where $P_{bf}$ denotes the collision probability of the 802.11bf AP, $L+1$ is the maximum retry limit. $W_j^{bf}$ is the contention window size of 802.11bf, which is $W_j^{bf}= 2^j CW_{\min}^{bf}$, where $CW_{\min}^{bf}$ is the minimum contention size window of 802.11bf. Unlike Bianchi's model, we have $P_f$ as the freezing probability, which captures the phenomenon that an AP in a backoff stage freezes its backoff counter when it senses the channel busy \cite{Ekici}. Similarly, the expression for $\tau_{ax}$ is given as
\begin{align}\label{coll1_prob}
    \tau_{ax}=\frac{(1-P_{ax}^{L+1})/(1-P_{ax})}{\left(\sum_{j=0}^{L}\left[1+\frac{1}{1-P_f} \sum_{k=1}^{W_j^{ax}-1} \frac{W_j^{ax}-k}{W_j^{ax}}\right] P_{ax}^j\right)},
\end{align}
where $P_{ax}$ denotes the collision probability of the 802.11ax AP. $W_j^{ax}$ is the contention window size of 802.11ax. 

\begin{figure*}[] 
	\setcounter{equation}{7}
	\begin{equation}
     \begin{aligned}\label{t_final}
         &T_m=(1-P_{t,ax})(1-P_{t,bf})\sigma + P_{t,ax}P_{s,ax}(1-P_{t,bf})T_{s,ax}+P_{t,bf}P_{s,bf}(1-P_{t,ax})T_{s,bf}+P_{t,ax}(1-P_{s,ax})(1-P_{t,bf})T_{c,ax}\\
         &+P_{t,bf}(1-P_{s,bf})(1-P_{t,ax})T_{c,bf}+\overline{T_{c}}(P_{t,ax}P_{s,ax}P_{t,bf}P_{s,bf}+P_{t,ax}P_{s,ax}P_{t,bf}(1-P_{s,bf})+P_{t,ax}(1-P_{s,ax})P_{t,bf}P_{s,bf}\\
         &+P_{t,ax}(1-P_{s,ax})P_{t,bf}(1-P_{s,bf}))
     \end{aligned}
\end{equation}
\hrule
\end{figure*}
\setcounter{equation}{2}
Further, the $\tau_{bf}$ and $\tau_{ax}$ are functions of $P_{bf}$ and $P_{ax}$, respectively. The probability of collision occurs when at least two APs out of $N_{bf}$ 802.11bf APs and $N_{ax}$ 802.11ax APs transmit simultaneously in the same time slot, i.e., 
\begin{align}\label{coll_prob}
    P_{bf}&=1-(1-\tau_{bf})^{N_{bf}-1}(1-\tau_{ax})^{N_{ax}}\\
    P_{ax}&=1-(1-\tau_{ax})^{N_{ax}-1}(1-\tau_{bf})^{N_{bf}}.
\end{align}

Let $P_{t,bf}$ and $P_{t,ax}$ denote the probability that at least one of the $N_{bf}$ 802.11bf APs and one of the $N_{ax}$ 802.11ax APs attempts to transmit in a given time slot. These can be expressed as,
\begin{align}\label{coll1_prob}
    P_{t,bf}&=1-(1-\tau_{bf})^{N_{bf}}\nonumber\\
    P_{t,ax}&=1-(1-\tau_{ax})^{N_{ax}}.
\end{align}

Now, let $P_{s,bf}$ and $P_{s,ax}$ denote the probability of successful transmission occurring on the channel by 802.11bf or 802.11ax under the condition that at least one of the AP (either 802.11bf or ax) transmits. These can be written as
\begin{align}\label{succ_prob}
    P_{s,bf}&=\frac{N_{bf}\tau_{bf}(1-\tau_{bf})^{N_{bf}}}{P_{t,bf}}\nonumber\\
    P_{s,ax}&=\frac{N_{ax}\tau_{ax}(1-\tau_{ax})^{N_{ax}}}{P_{t,ax}}.
\end{align}
We will use the probabilities in Equations~\eqref{coll_prob} to~\eqref{succ_prob} for subsequent throughput and mean delay derivations in the next sections.
\subsection{Throughput Analysis of 802.11ax}
The normalized throughput or MAC efficiency of the 802.11ax network is defined as the fraction of successful packet transmissions over the mean time of all possible events. It is a measure of how effectively the medium is used for successful data transmissions, and can be expressed as
\begin{equation}\label{throug_final}
S=\frac{P_{t,ax}P_{s,ax}(1-P_{t,bf})T_{f,ax}}{T_m},
\end{equation}
where $T_{f,ax}$ is the total frame duration of 802.11ax. $T_m$ is the average time of all possible events, such as idle, collision or successful transmission events, and is given in Equation~\eqref{t_final} (on top of next page).
Where, $\sigma$ is the duration of an idle slot, $T_{s,ax} (T_{s,bf})$ is the successful transmission duration of 802.11ax (802.11bf), and $T_{s,ax} (T_{s,bf})$ is the collision duration of 802.11ax (802.11bf), and $\overline{T_{c}}$ is the cross-technology collision duration, i.e., $\max(T_{c,ax}, T_{c,bf})$. The expressions for all these parameters are explained subsequently.

\subsubsection{Expression for $T_{f,x}$}
Let the duration of the frame, $T_{f,x}$, depends on the technology and is given by:
\setcounter{equation}{8}
\begin{equation}\label{eq_23}
T_{f,x}=\begin{cases}
   \begin{split} PHY&_{header}+\frac{MAC_x}{R}, \text{if $x$ is 802.11ax AP} \end{split}\\
   \begin{split} T_\textit{CFP}& , \text{if $x$ is 802.11bf AP,} \end{split} 
   \end{cases}
\end{equation}
Here, $MAC_x = MAC_{header}+MSDU$, i.e., duration of MAC header and MAC service data unit size (MSDU) and $R$ is the transmission data rate. For 802.11bf, the frame duration is equal to the total Contention-Free Period (CFP), i.e., the total duration of all three phases, as shown in Figure \ref{fig:System_model1}, i.e.,
\begin{equation}
\begin{aligned}
\label{report_mid}
T_\textit{CFP} =& \underbrace{T_{Polling} + SIFS + T_{CTS}}_{\text{Polling Phase}} + SIFS \\
& + \underbrace{ N \times \big[T_{NDPA} + SIFS + T_{NDP}\big]}_{\text{Sounding Phase}} + SIFS \\
& + \underbrace{ N \times T_{Reporting}}_{\text{Reporting Phase}},
\end{aligned}
\end{equation}
where $T_{Polling}$, $T_{CTS}$, $T_{NDPA}$, and $T_{NDP}$ are the transmission duration of polling, CTS-to-self, NDPA packet, and NDP frames, respectively. These are all 802.11bf standard specific parameters \cite{TGbf_nist}. Further, $N$ represent the number of STA participants in the sensing procedure.

The duration of the reporting frame, i.e. $T_{Reporting}$ (last term in Equation \eqref{report_mid}), is mainly dependent on the CSI report size of each participating STA. Specifically in the 802.11bf standard~\cite{bf-draft}, the CSI size is given by the following formula:
\begin{equation}
\begin{aligned}
\text{CSI Report Size} =& \left\lceil 1.5 * N_{\text{tx}} * N_{\text{rx}} \right\rceil + \frac{N_{\text{tx}} * N_{\text{rx}} * N_b * N_{\text{sc}}}{4}&\\ + 2 * N_{\text{tx}}
\end{aligned}
\label{csi_size}
\end{equation}
The CSI size depends on the number of transmitting antennas (\(N_{\text{tx}}\)), receiving antennas (\(N_{\text{rx}}\)), subcarriers (\(N_{\text{sc}}\)), and the number of bits used for quantizing each CSI value (\(N_b\)). Now, the duration of $T_{Reporting}$ can be determined by dividing the size of the CSI report by the transmission rate $(R)$, i.e.,
\begin{equation}
T_{Reporting}=\frac{\text{CSI Report Size}}{R}.
\end{equation}
The effect of the number of transmitting and receiving antenna elements and corresponding CSI report size on the overall sensing performance is examined in Section V-E.

\subsubsection{Expression for $T_{s,x}$}
The duration of a successful transmission, $T_{s,x}$, is:
\begin{equation}\label{eq_24}
T_{s,x}=\begin{cases}
    \begin{split} T_{f,x}&+DIFS+SIFS+PHY_{header}+\frac{ACK}{R_{min}}, \\&\text{if $x$ is 802.11ax AP},  \end{split}\\
    \begin{split} T_{f,x}&+DIFS&\text{if $x$ is 802.11bf AP},  \end{split}
    \end{cases}
\end{equation}
\subsubsection{Expression for $T_{s,x}$}
The collision duration, $T_{c,x}$, in Equation~\eqref{t_final} is given by 
\begin{equation}\label{eq_242}
T_{c,x}=\begin{cases}
    \begin{split} T_{f,x}&+DIFS, \text{if $x$ is 802.11ax AP} \end{split}\\
    \begin{split} T_{f,x}&+DIFS, \text{if $x$ is 802.11bf AP.} \end{split}
    \end{cases}
\end{equation}

Substituting the Equations \eqref{eq_24}-\eqref{eq_242} in Equation \eqref{t_final}, we obtain the final expression for $T_m$. Further, using the expressions of $T_m$ and $T_{f,x}$ (from Equation \eqref{eq_23}) in the Equation \eqref{throug_final}, we obtain the final normalized throughput expression for the 802.11ax network.

\subsection{Sensing Delay Analysis of 802.11bf}
\label{sen_delay_anal_bf}
\begin{figure*}[!ht]
    \centering
    \subfigure[{802.11bf Latency Distribution}]{
    \hspace{-0.9mm}
    \includegraphics[width=0.46\textwidth,trim = 0mm 0mm 0mm 0mm,clip]{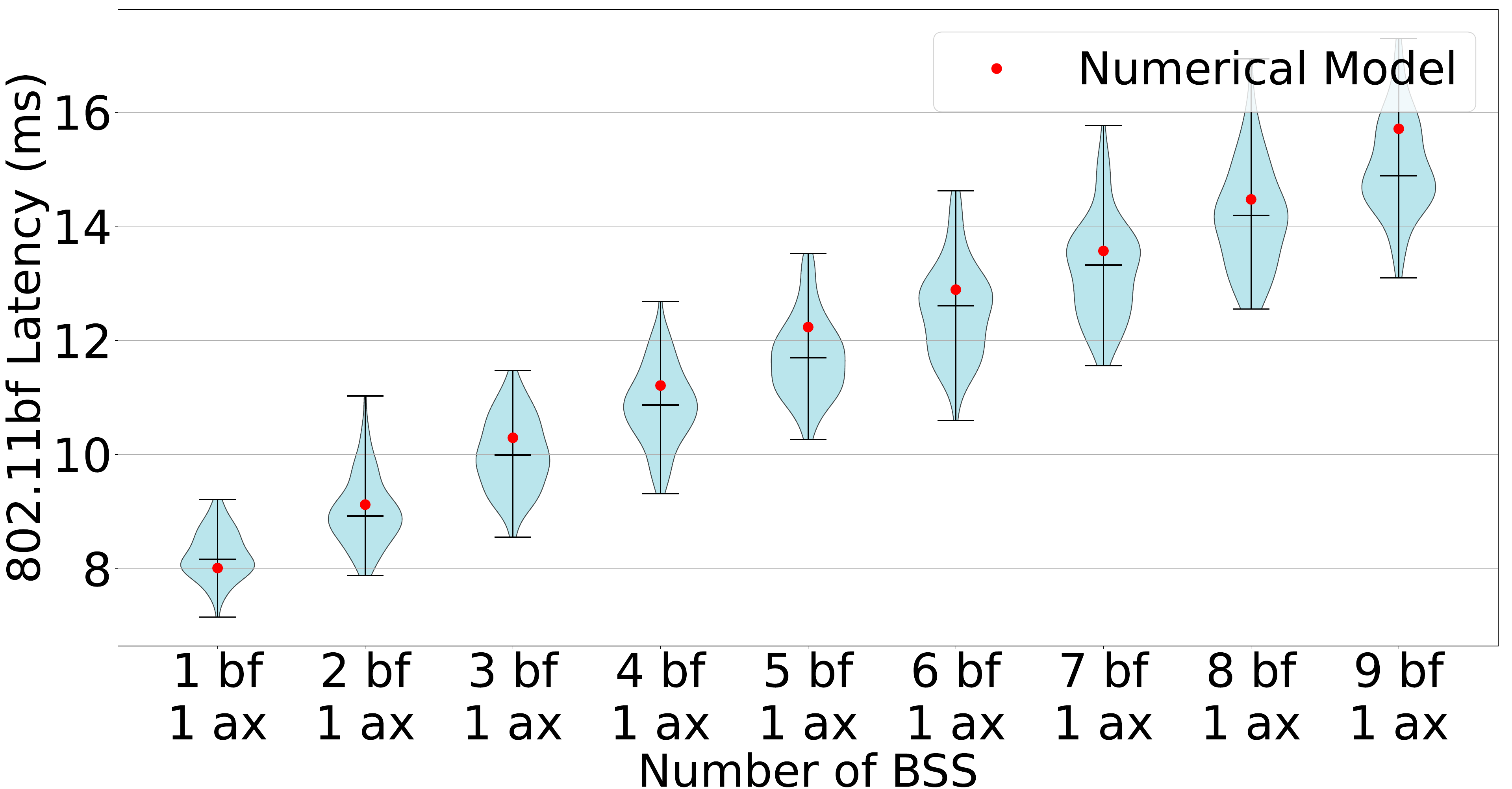}
    \label{TSUS_1_1}
    }
    \subfigure[{802.11ax Throughput Distribution}]{
\hspace{-1mm}\includegraphics[width=0.47\textwidth,trim =0mm 0mm 0mm 0mm,clip]{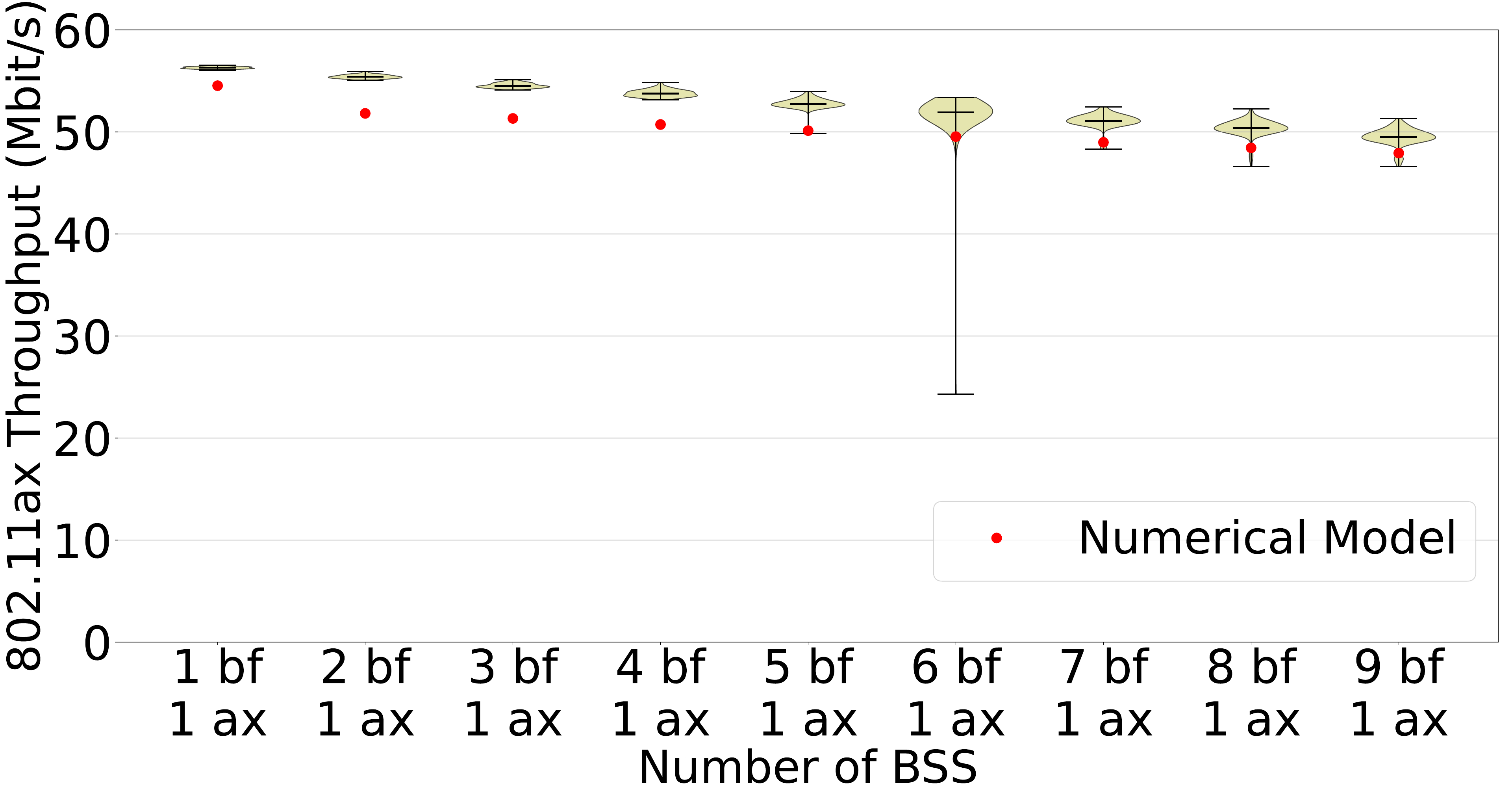}
    \label{TSUS_1_2}
    }
    \caption{The effect of increasing the number of 802.11bf AP on sensing latency and 802.11ax throughput.}
    \label{fig:scenario-1}
  \end{figure*}

  \begin{figure*}[!ht]
  \centering
    \subfigure[{802.11bf Latency Distribution}]{
    \includegraphics[width=0.46 \textwidth,trim =0mm 0mm 0mm 0mm,clip]{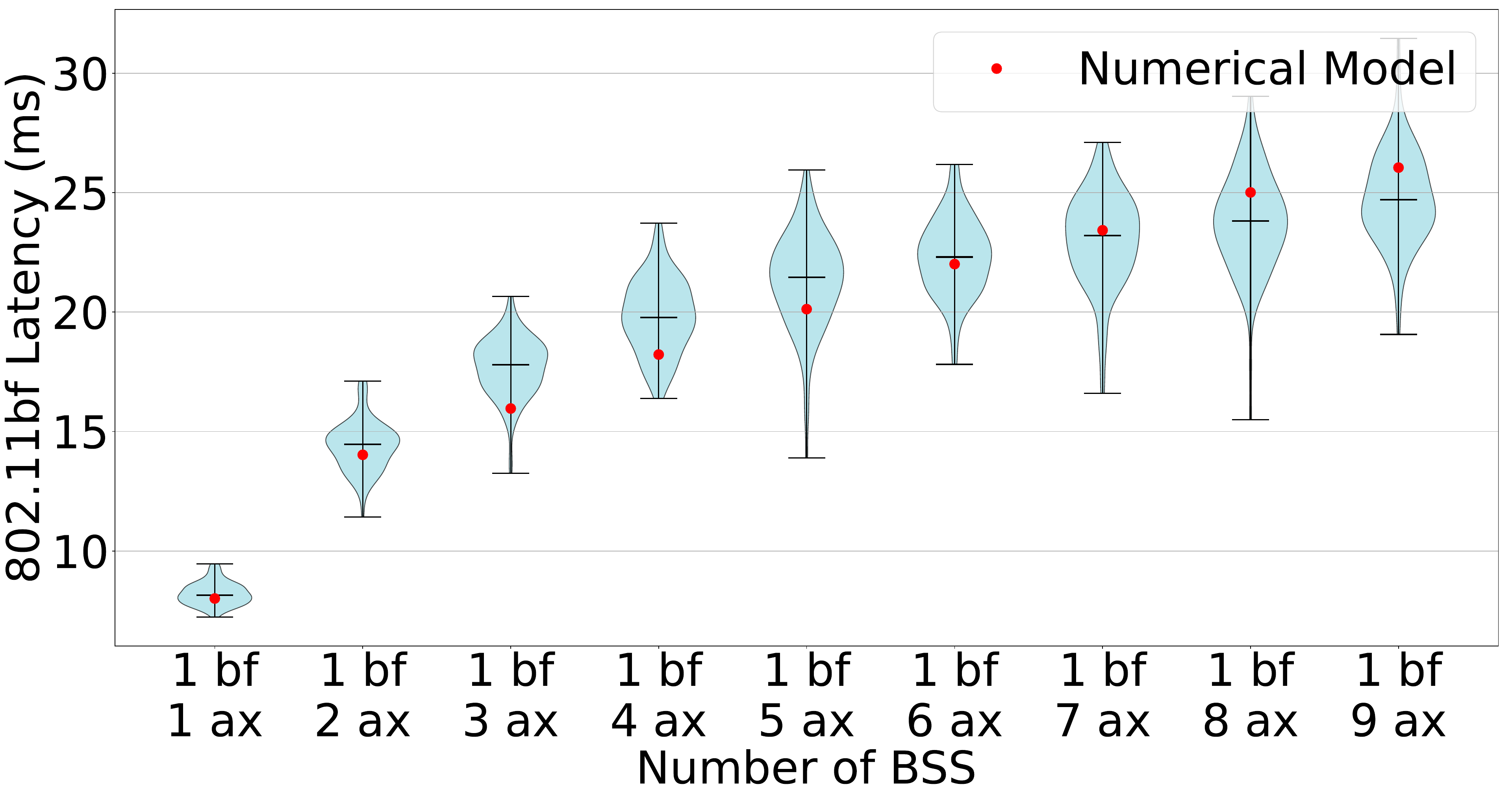}
    \label{TSUS_1_3}
    }
    \subfigure[{802.11ax Throughput Distribution}]{
    \includegraphics[width=0.47\textwidth,clip]{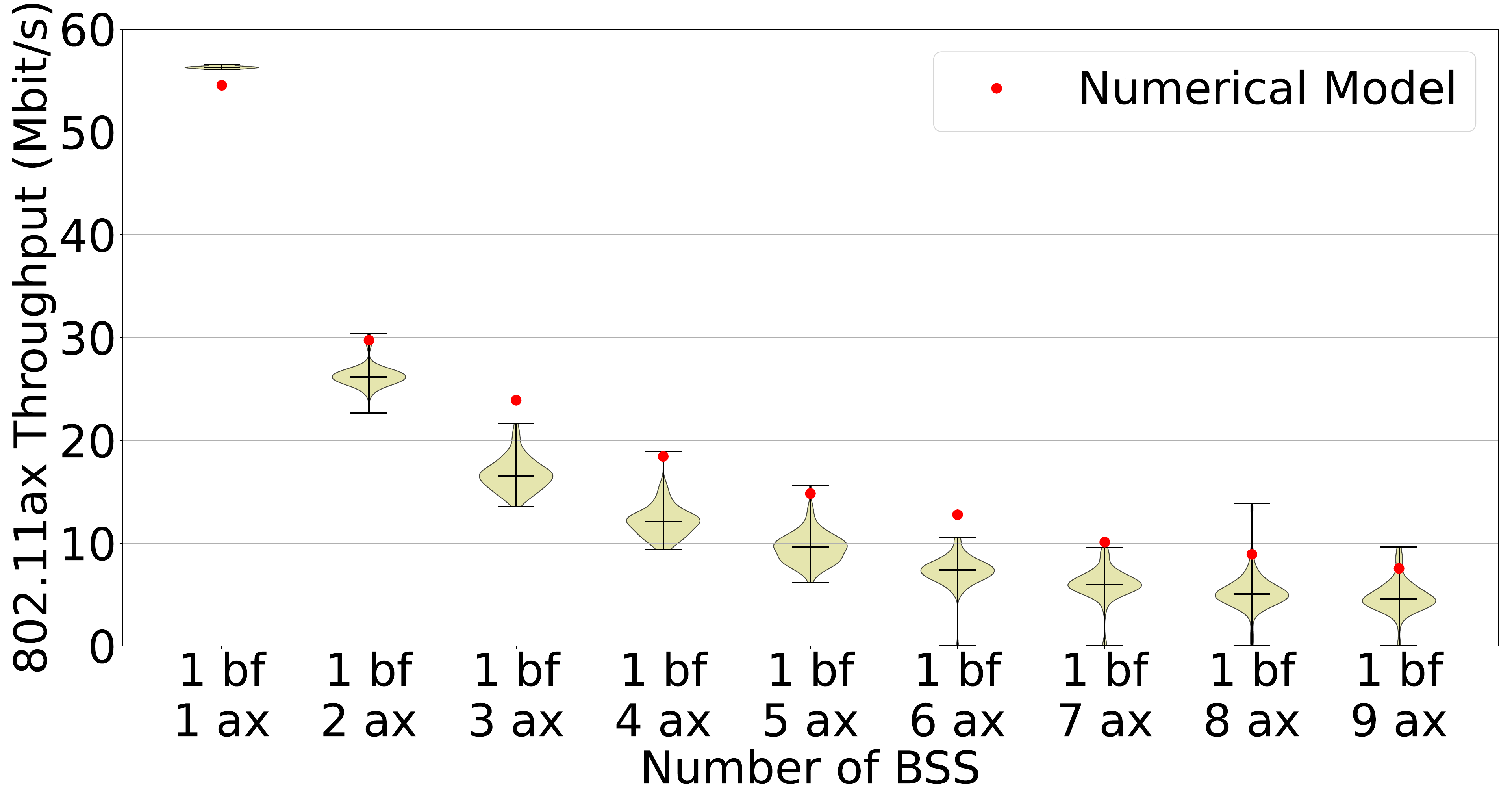}
    \label{TSUS_1_4}
    }
    \caption{The effect of increasing the number of 802.11ax AP on sensing latency and 802.11ax throughput.}
	\label{fig:scenario-2}
\end{figure*}
In this subsection, we derive the expression for the sensing delay of 802.11bf network. The 802.11bf APs perform CSMA/CA to access the channel in presence of IEEE 802.11 APs. Once the 802.11bf AP has accessed the channel, it starts the sensing measurement procedure, which consists of four phases as discussed in Section \ref{sec:primer}. The overall sensing latency at one measurement instance is defined from the time the AP starts contending for the channel till the end of the reporting phase. The sensing latency in 802.11bf is mostly attributed to the channel access delay, primarily caused by collisions and backoff procedures, as all phases outlined in the standard are deterministic. 

Let the 802.11bf AP gets access to the channel in its $(i+1)$-th attempt. The average channel access delay after experiencing $i$ collisions can be written as follows:
\begin{equation} 
T^{(i)}=T_\textit{CFP}+i*T_{c,bf}+\sum_{j=0}^i \overline{W_j} \mathcal{F},
\end{equation}
where $T_\textit{CFP}$ is the CFP period as in Equation \eqref{report_mid}, $T_{c,bf}$ is the collision duration expressed in Equation \eqref{eq_242}, $\overline{W_j}$ is the average number of backoff slots in stage $j$, i.e., $\overline{W_j}=\frac{W_j-1}{2}$, and $\mathcal{F}$ is the average slot duration that the AP waits in a stage. 

Let $P^{(i)}_{\text {suc}}$ denote the probability of the AP getting access to the channel in $i+1$ attempt, then, from Equation \eqref{coll_prob}, $P^{(i)}_{\text {suc}}=(1-P_{bf})P_{bf}^i$. Conditioning on the probability of success, the overall channel access delay of AP is
\begin{equation} \label{overal_delay}
T=\frac{1}{1-P_{bf}^{L+1}} \sum_{i=0}^L P_{\text {suc}}^{(i)}\left[ T_\textit{CFP}+i*T_{c,bf}+\left(\sum_{j=0}^i \overline{W_j} \mathcal{F}\right)\right],
\end{equation} 
where $L$ is the maximum backoff stage. 
From Equation \eqref{overal_delay}, the expression for the overall channel access delay of an AP depends on the average duration it stays in each backoff stage before the backoff decrement occurs, i.e. $\mathcal{F}$. To calculate the $\mathcal{F}$, we need to consider when we enter the backoff states, which are either from a previous backoff state or from a transmission state. As $\tau_{bf}$ is the transmission probability of the AP, $1-\tau_{bf}$ is the probability of remaining in backoff state. Therefore, the average slot duration $\mathcal{F}$ is
\begin{equation}\label{f_final}
\mathcal{F}=(1-\tau_{bf}) F_b+\tau_{bf} F_t,
\end{equation}
where $F_b$ is the average time duration AP stays in the current backoff state while entering from a previous backoff state, and $F_t$ is the average time duration the AP stays in the backoff state following an attempt, either successful or collision. Let $d_I$, $d_S$, and $d_C$ denote the duration that the AP stays in the backoff state when the channel status at the beginning of that slot is idle, involved in a successful event, or experiencing a collision, respectively. Detailed calculations for these durations are provided in the Appendix.

Let $P_d$ be the backoff decrement probability, which depends on the probability of the channel being idle. In that case, conditioning on $P_d$, the average duration AP will stay in the current stage $F_b$ is
\begin{equation}
    F_b=\frac{P_{i} d_I+P_{s} d_S+P_{c}d_C}{P_d}, 
\end{equation}
where $P_{i} = 1 - P_{bf}$, the expression of $P_{bf}$ is defined in Equation \eqref{coll_prob}, which represents the probability that no transmission occurred in the slot. $P_{s}$ denote the probability that when AP enters to backoff stage and finds the channel busy due to successful transmission (either from another 802.11bf AP or 802.11ax APs). The expressions for $P_{s}$ is
\begin{equation}
\begin{aligned}
 P_{s}=&\binom{N_{bf}-1}{1} \tau_{bf}(1-\tau_{bf})^{N_{bf}-2} (1-\tau_{ax})^{N_{ax}} \\& +\binom{N_{ax}}{1}\tau_{ax}(1-\tau_{ax})^{N_{ax}-1}.
  \end{aligned}
\end{equation}
Further, $P_{c}$ denotes the probability that the previous transmission ended in collision, i.e., $P_{c} = 1- P_{s}-P_{i}$.

Now, we derive the expression for $F_t$. After each attempt, either successful or collision, the AP selects a new backoff counter. Let $\overline{CW}$ be the average backoff window size over all stages. Then, the probability that the AP will content for the channel is $1 / \overline{C W}$ and the probability that it will go to one of the backoff stages is $1-1 / \overline{C W}$. In this case, the average slot duration when the AP enters from a transmission state:
\begin{equation}
F_t=\frac{(\overline{C W}-1)(P_{i} d_I+P_{s} d_S+P_{c}d_C)}{\overline{C W}},
\end{equation}
where the expression for $\overline{C W}$ is
\begin{equation}
\overline{C W}=\frac{1}{1-P_{bf}^{L+1}}\sum_{i=0}^L(1-P_{bf}) P_{bf}^i W_i.
\end{equation}
Substituting the expressions for $F_b$ and $F_t$ in Equation \eqref{f_final}, we obtain the average channel access delay of 802.11bf AP.
\begin{figure*}[ht]
  \centering
    \includegraphics[width=0.98\textwidth,trim = 0mm 0mm 0mm 0mm,clip]{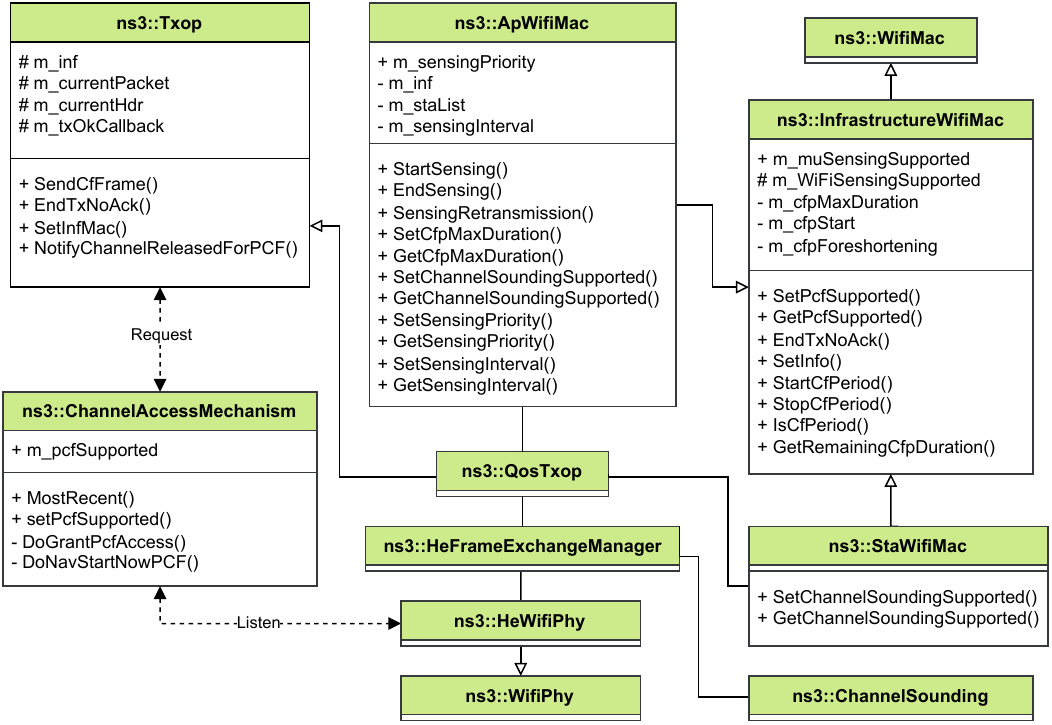}
    \centering
    \caption{UML diagram of the \mbox{ns-3} 802.11bf implementation.}
	\label{fig:UML_class_diagram}
\end{figure*}
\subsubsection{Sensing Interval}
\label{subsec::sensing_interval}
Up to this point, we have assumed that the application continuously requests the sensing procedure. However, to make the model more realistic, we assume that the application requests sensing according to a Poisson arrival process. In this case, the probability that doesn't have sensing request is equal to 
\begin{equation}
\label{P_0_poi}
P_0=1-\frac{\lambda}{\bar{\mu}}.
\end{equation}
In this study, we assume that 802.11ax devices have saturated traffic, while 802.11bf devices operate under non-saturated conditions with sensing requests following a Poisson arrival process. Under these conditions, we can calculate the mean channel access delay when $i$ number of 802.11bf APs are contending for channel access with Poisson-based sensing requests, while all 802.11ax APs are operating in a saturated state and continuously attempt to access the channel.
So, the total number of contending APs at any given time is $N_{ax}$+ $n_{bf}^i$, where the number bf AP which are contending $n_{bf}^i$ follows a binomial distribution with probability:
\begin{equation}
\begin{aligned}
&n_{bf}^i=\binom{N_{\text {bf }}}{i}\left(1-P_0\right)^{i} P_0^{N_{bf}-i}.
\end{aligned}
\end{equation}
Therefore the expected channel access delay can be calculated as follows given that $N_{ax}$ 80.11ax and $n_{bf}^i$ 80.211bf APs are contending :
\begin{equation}
E[T]=\sum_{i=1}^{N_{bf}} n_{bf}^{i}\times {T^{(i)}}.
\end{equation}
The average service rate is 
\begin{equation}
\label{mu-mid}
\bar{\mu}=\frac{1}{E[T]}.
\end{equation}
Equations \eqref{P_0_poi} and \eqref{mu-mid} are coupled with each other and can be solved iteratively using the numerical technique\cite{Ekici, 840210}.

In what follows, we first introduce our ns-3 implementation of 802.11bf, and then use it to analyze the coexistence of 802.11bf and 802.11ax networks.
\section{The \mbox{ns-3} 802.11bf implementation}
In this Section, we provide an overview of our 802.11bf module implementations in \mbox{ns-3}. We emphasize that this is the first implementation of 802.11bf within \mbox{ns-3}. Figure~\ref{fig:UML_class_diagram} illustrates the UML diagram of our implementation, highlighting the modifications, newly added attributes, and functions integrated into the \mbox{Wi-Fi} module classes to accurately model the sensing behavior of 802.11bf. Our implementation is open source and available at \cite{bf_ns3}, allowing the research community to reproduce our work and extend it in future research.

\mbox{ns-3}~\cite{henderson2008network} is an open-source discrete-event network simulator widely used for modeling and analyzing communication systems. It is designed in a modular structure, where each module represents the behavior of a specific technology.  The \mbox{Wi-Fi} module in \mbox{ns-3} provides detailed implementations of IEEE 802.11 standards. It supports various IEEE 802.11 standards, including 802.11a/b/g/n/ac/ax/be \cite{pk_lcn,navid_icmlcn2024,zorziTCOM_2023}. 

We implemented 802.11bf within \mbox{ns-3.40}, by extending the existing \mbox{Wi-Fi} module, which leverages 802.11ax’s EDCA and Multi-User OFDMA (MU-OFDMA) features. These features provide efficient channel access and allow for simultaneous uplink transmissions by multiple users. We adapted \mbox{ns-3}, most importantly the TB-based sensing measurement procedure (\textit{cf}. Section \ref{subsec:TB_config}), to comply with 802.11bf’s \mbox{Wi-Fi} sensing requirements. To mimic the 802.11bf operation, we implemented a similar PCF mechanism layered on top of EDCA. This integration allows the AP to initiate sensing sessions by polling 
 the associated STAs for CSI information. This polling process begins with a trigger frame sent by the AP, and associated STAs respond with CTS-to-Self frames, signaling their readiness for sensing. Modifications to the \texttt{ns3::InfrastructureWifiMac} class were made to enable these contention-free periods, which are critical for preventing interference during CSI measurement.

We implemented three key phases: polling, NDPA sounding, and reporting. We incorporated the NDPA, NDP, and reporting frames based on the implementation by Zhang \textit{et al}. in \cite{Zhang-ns3} for 802.11ax channel sounding. We also modified the \texttt{ns3::MultiUserScheduler} to manage Resource Unit (RU) allocation, allowing MU OFDMA transmissions that enable simultaneous reception of CTS-to-self frames by multiple STAs.
In the polling phase, the AP initiates the sensing session by sending polling trigger frames to associated STAs, which respond with CTS-to-Self frames, indicating their readiness. During the NDPA sounding phase, the AP transmits an NDPA frame followed by an NDP to each STA. The NDP enables STAs to measure the channel and obtain CSI based on randomized channel conditions, simulating real-world variability. 
In the reporting phase, as shown in \mbox{Figure \ref{fig:System_model1}}, STAs send their CSI back to the AP using Beamforming Report Poll (BFRP) frames. Our implementation supports both immediate and delayed reporting methods to provide flexibility under different network conditions. For scenarios involving multiple STAs, we used MU-OFDMA in the uplink direction as well, permitting concurrent transmission of CSI reports. The \texttt{ns3::MultiUserScheduler} was further modified to handle RU allocation for these uplink transmissions during the reporting phase.
\section{Performance Evaluation}
\label{sec:PerformanceEvaluation}
In this section, we present simulation results obtained using our own  \mbox{ns-3} module for 802.11bf, specifically designed to evaluate its coexistence with 802.11ax networks and validated against our analytical expressions obtained in Section III. 
\subsection{Simulation Setup}
We use two different environment layouts for our studies as follows. First, we consider a generic $100~m\times100~m$, \textit{layout I}, where the 802.11bf and 802.11ax APs are randomly located. Each AP, regardless of whether it uses 802.11bf or 802.11ax, is connected to two STAs, and this network setup is called a one basic service set (BSS) \cite{bertsekas2021data}. Second, we adopt a realistic indoor office deployment, namely \textit{layout II}, based on the 3GPP TR 38.901 standard \cite{3GPP2}. This layout incorporates detailed channel characteristics and structural features such as corridors, open spaces, and partitioned work areas.

The simulations investigate the impact of key network parameters in five scenarios: (i) varying the number of APs for both 802.11bf and 802.11ax, (ii) assigning different access categories to 802.11bf sensing APs, (iii) varying different sensing intervals, (iv) changing the number of antenna elements used for sensing in 802.11bf APs, and (v) varying the channel bandwidth. These scenarios are designed to capture how network configurations influence the coexistence of sensing and communication functionalities. The first four scenarios are evaluated within the generic layout, while the fifth scenario is specific to the indoor office deployment.

The analysis focuses on two performance metrics: the sensing delay of 802.11bf and the aggregated downlink throughput of 802.11ax networks. Unless otherwise stated, the simulation parameters are summarized in Table~\ref{tab:parameters}.

\begin{table}[h!]
	\caption{Simulation Parameters}
	\begin{tabular}{|p{3cm}|p{5cm}|}
		\hline
		Layout& Simple 100×100 meters layout and office layout from 3GPP TR 38.901
		\\
		\hline
		Carrier frequency& 5 GHz\\
        \hline
		Sensing interval& 100~ms\\
		\hline
		Transmit power& 23 dBm 
        \\
		\hline
		Channel bandwidth &20~MHz, 80~MHz, 160~MHz
		\\
		\hline
	    Frame Aggregation&  
	        A-MPDU 64 Frame
		\\
		\hline
		Simulation duration& {10~s} 
		\\
		\hline
		802.11ax Traffic& Downlink saturated best-effort traffic with a payload of 1474 bytes per packet
		\\
		\hline
	
		  Sensing threshold&{ -62 dBm} 
		\\
		\hline
		 	$PHY_{header}$& 20 $\mu$s 
		 \\
		\hline
		$DIFS$& 34 $\mu$s 
        \\
		\hline
        $SIFS$& 16 $\mu$s 
		\\
		\hline
			No. of seeds&{50 iterations for each network size}
		\\
		\hline
	\end{tabular}
	\label{tab:parameters}
\end{table}
\subsection{Effect of Varying 802.11bf and 802.11ax APs}
In this case study, we consider \textit{layout I} with two separate scenarios. In the first scenario, we vary the number of 802.11bf APs while keeping a single  802.11ax AP. This approach allows us to more clearly distinguish and understand the impact of introducing  802.11bf on the performance of  802.11ax. Similarly, in the second scenario, we reverse the roles: we fix one 802.11bf AP and vary the number of 802.11ax APs. For each variation in both scenarios, we run the simulation with 50 different seeds to capture a wide range of environmental conditions. We then present the simulation results using violin plots to visualize the underlying distribution, where the width reflects data density, and we use markers to represent numerical results. 

In the first scenario, illustrated in Figure~\ref{fig:scenario-1}, increasing the number of 802.11bf APs leads to a noticeable rise in sensing latency. As shown in Figure~\ref{TSUS_1_1}, the median latency at one 802.11bf AP is approximately 8~ms, but increases to about 15~ms when nine 802.11bf APs are deployed. This growth in latency primarily stems from channel access delays and the corresponding polling ($T_{polling}$) and reporting ($T_{Reporting}$) phases, as detailed in Section~\ref{sen_delay_anal_bf}.

In Figure~\ref{TSUS_1_2}, we see that introducing additional 802.11bf APs exerts only a minor influence on 802.11ax throughput, which decreases slightly from 56~Mbps to 50~Mbps. The limited impact arises from the fact that 802.11bf’s sensing operations briefly occupy the channel. Moreover, in our simulations, the sensing interval is set to $100~ms$, meaning the sensing activity is infrequent compared to the total simulation duration. The effect of this configuration is examined later in Section\ref{subsection:sensing-intervals}. Furthermore, this plot confirms that the analytical and simulation models closely match in terms of the 802.11bf sensing latency and the aggregated 802.11ax downlink throughput.

In the second scenario, as shown in Figure~\ref{TSUS_1_3}, increasing the number of 802.11ax APs leads to a significant rise in the sensing latency of the 802.11bf network. For instance, with one 802.11ax AP, the latency of the  802.11bf network is approximately 8~ms, but it increases to 26~ms when the number of 802.11ax APs reaches nine. The primary cause of this latency growth is the heightened contention for channel access, which delays the 802.11bf AP from securing the channel for its sensing operations. Moreover, the 802.11ax APs tend to capture the channel for longer periods due to saturated traffic and frame aggregations, causing the sensing operations to experience longer waiting times before they can commence. 
In Figure~\ref{TSUS_1_4}, we observe that the mean throughput of the 802.11ax networks decreases from approximately 56~Mbps with one 802.11ax AP to around 5~Mbps when there are nine 802.11ax APs. This reduction is expected due to increased competition and potential collisions among the 802.11ax networks themselves. Each AP  have to contend more frequently for channel access, leading to reduced available airtime and lower throughput per network.
Furthermore, the numerical and simulation results closely align under most network conditions. The slight discrepancies between the analytical and simulation outcomes stem from the fact that the numerical analysis employs averaged values for random parameters (e.g., contention window sizes), whereas the simulations inherently capture random variations in these parameters.

Finally, comparing Figure~\ref{TSUS_1_1} and Figure~\ref{TSUS_1_3}, we find that the sensing latency of 802.11bf is higher when coexisting with 802.11ax APs than when coexisting with additional 802.11bf APs. More specifically, in the presence of 802.11bf APs, the sensing delay increases linearly as a function of the number of APs. This behavior occurs because the overall sensing latency is primarily dictated by the polling ($T_{polling}$) and reporting ($T_{Reporting}$) phases, both of which scale linearly with the number of APs, as shown in Equation~\eqref{report_mid}.
In contrast, when 802.11bf coexists with a growing number of 802.11ax APs, the sensing latency increases nonlinearly, as shown in Equation~\eqref{overal_delay}. This is due to the dominant influence of the channel access delay. As 802.11ax APs aggressively capture the channel and maintain it for longer periods, the 802.11bf AP experiences significantly prolonged waiting times before it can perform its sensing operations.
Similar to the sensing delay, the aggregated throughput of 802.11ax APs suffers more severely when these APs coexist among themselves than when they coexist with 802.11bf APs
\subsection{Effect of Assigning Different Access category (AC) to Sensing Networks}
\begin{figure}[t]
\centering\includegraphics[width=0.43\textwidth,trim = 30mm 0mm 0mm 0mm,clip]{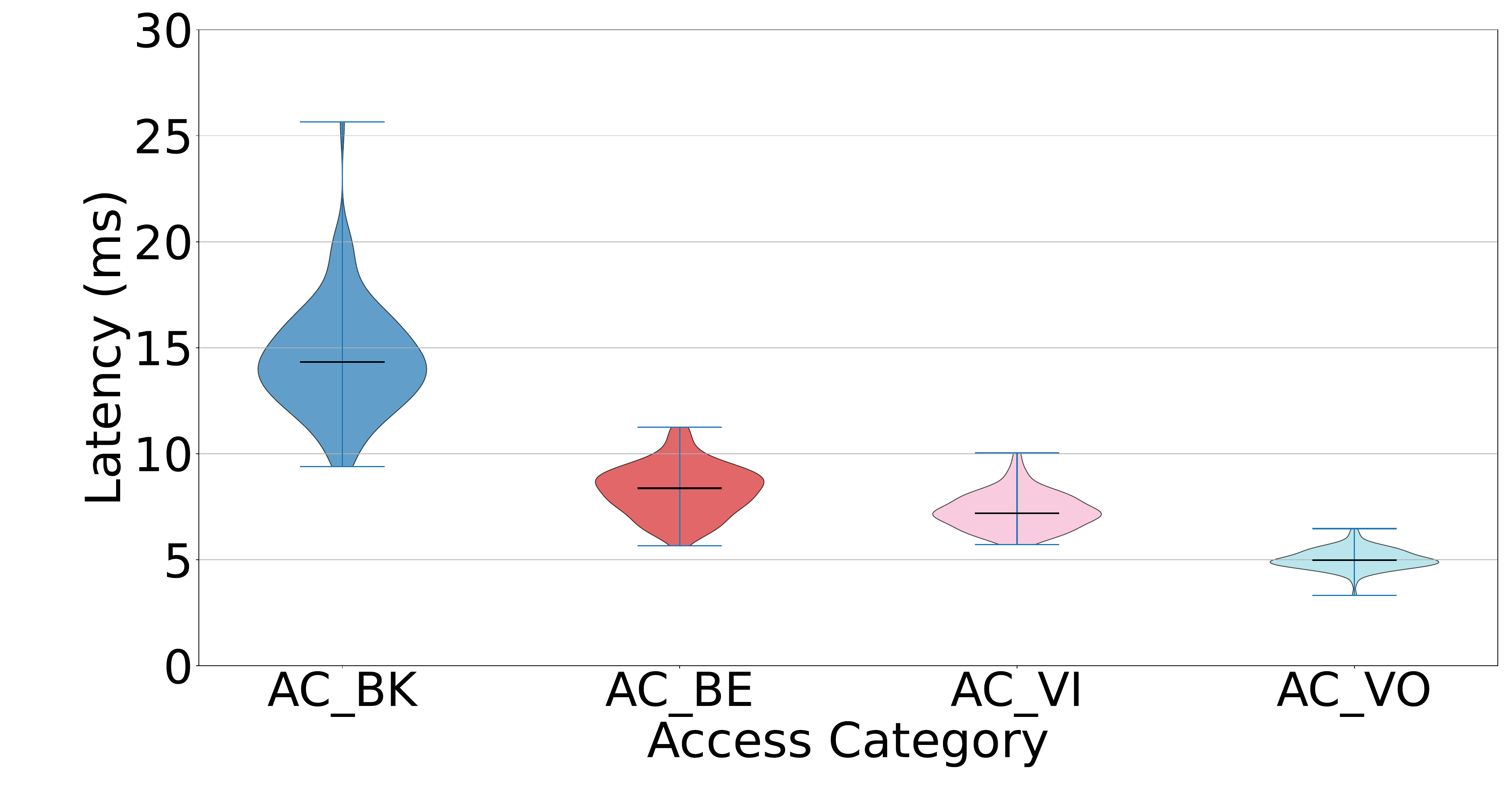}
	\caption{Effect of different access categories on the 802.11bf latency.}
	\label{fig:results_AC}
\end{figure}

The results presented in Figure~\ref{fig:results_AC} illustrate how varying the access categories (ACs) affects latency for a single 802.11bf sensing AP coexisting with an 802.11ax AP. We evaluate all ACs defined in Table \ref{tab:edca_parameters}—background (BK), best-effort (BE), video (VI), and voice (VO)—while keeping the 802.11ax AP fixed at AC\_BE.
We find that AC\_BK exhibits the highest median latency, approximately 15~ms, with substantial variability ranging from about 10~ms to 25~ms. This elevated latency and broad dispersion in the violin curve are due to AC\_BK's lower priority channel access parameters, particularly its larger contention window. In contrast, the VO access category achieves the lowest median latency of roughly 5~ms, with a narrow range of 4~ms to 7~ms, reflecting its higher priority and shorter contention periods (smaller CW and AIFS). Overall, these findings suggest that employing higher-priority ACs in 802.11bf, such as AC\_VI and AC\_VO, can effectively reduce latency in coexistence scenarios, thereby improving performance for time-sensitive sensing applications.

\subsection{Effect of Assigning Different Sensing Intervals}
\label{subsection:sensing-intervals}
\begin{figure*}[!ht]
  \centering
    \subfigure[{802.11bf Average Latency}]{
    \includegraphics[width=0.47\textwidth,clip]{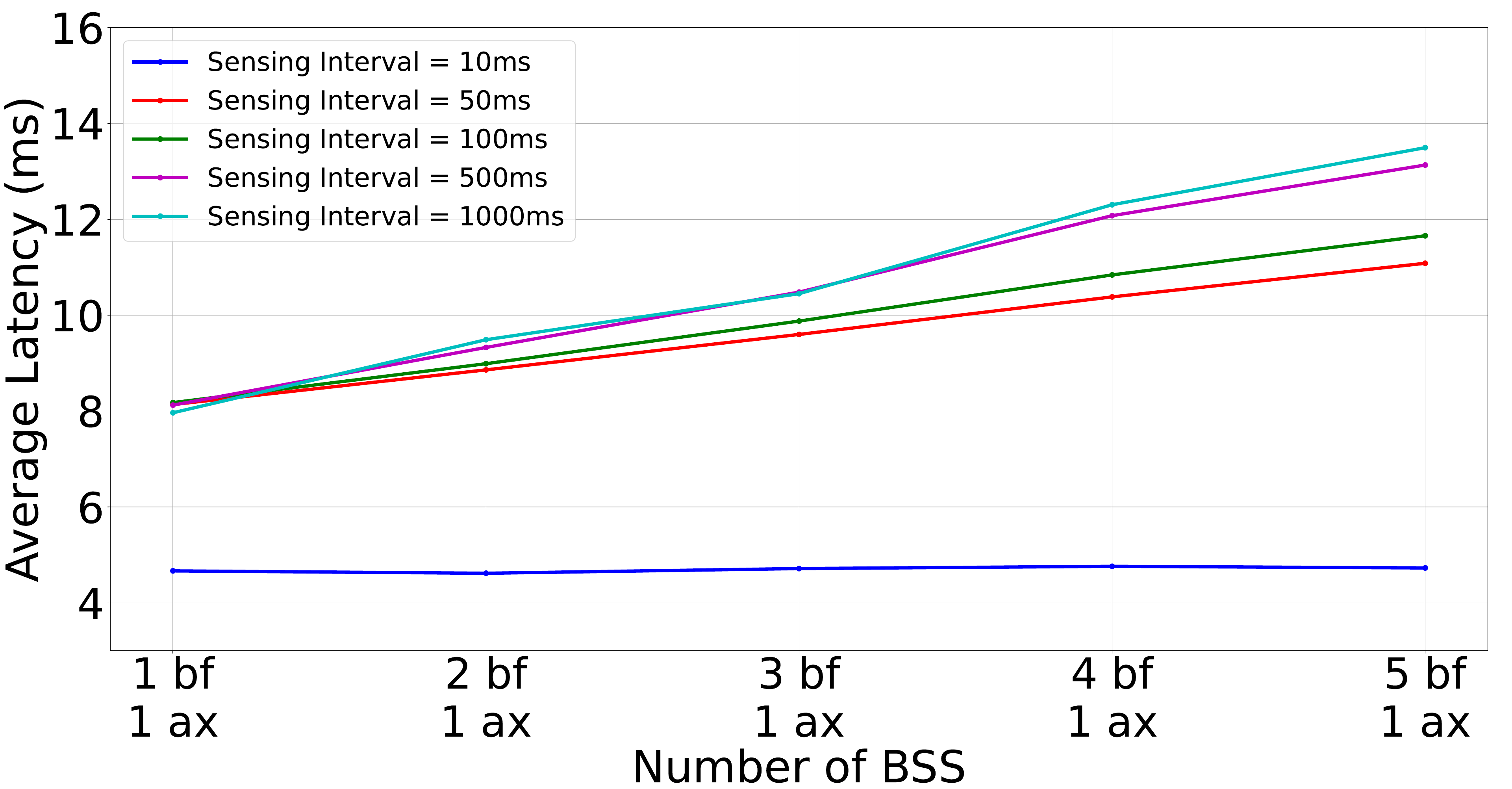}
    \label{TSUS_1_33}
    }
    \subfigure[{802.11ax Average Throughput}]{
    \includegraphics[width=0.47\textwidth,clip]{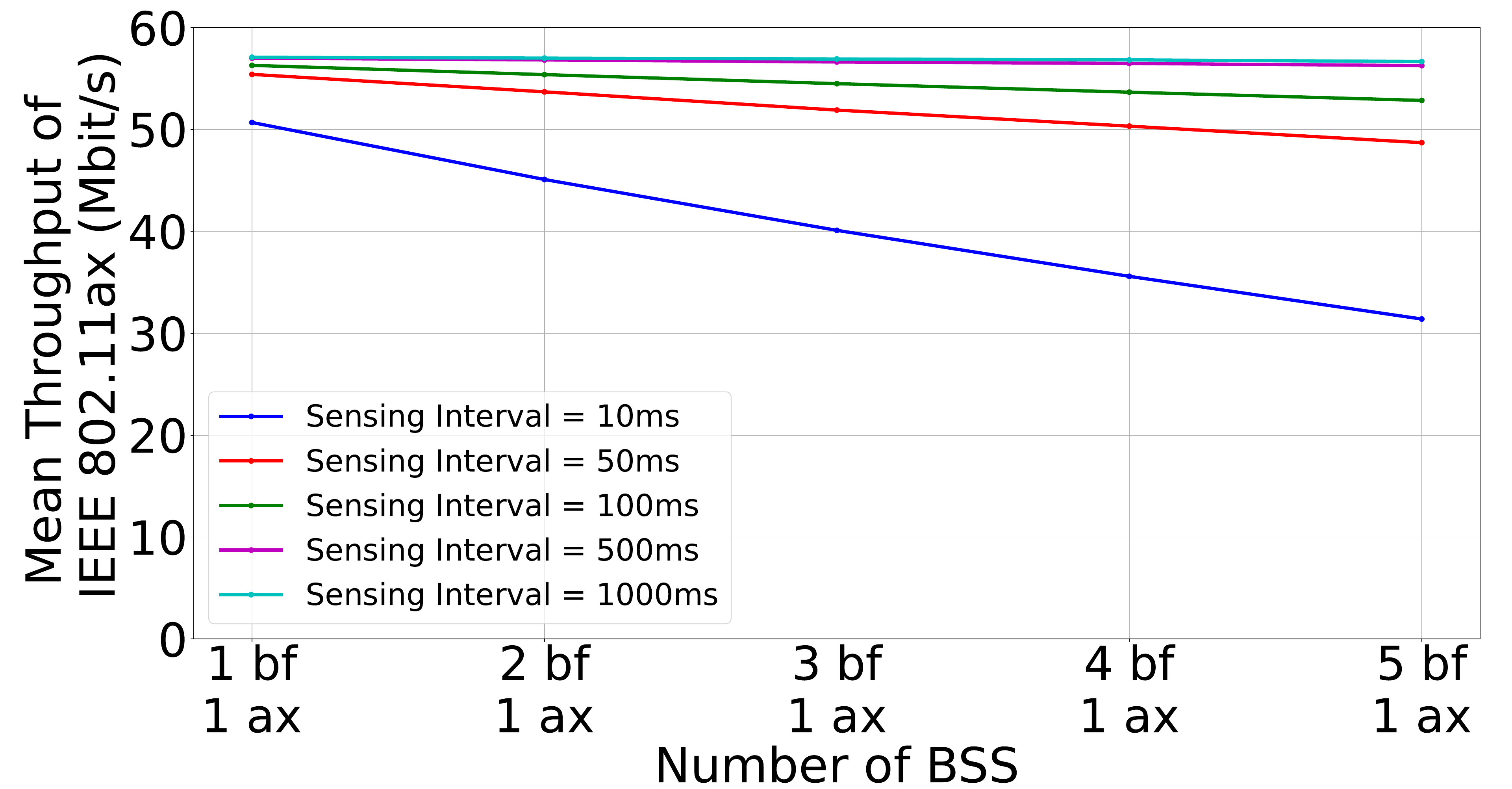}
    \label{TSUS_1_7}
    }
    \caption{The effect of different sensing intervals on sensing latency and 802.11ax throughput.}
	\label{fig:results_intervals}
\end{figure*}

In Figure~\ref{fig:results_intervals}, we show how varying sensing intervals influence overall coexistence performance. We consider five different intervals—$10~\text{ms}$, $50~\text{ms}$, $100~\text{ms}$, $500~\text{ms}$, and $1000~\text{ms}$—each corresponding to how often the application layer generates sensing requests. For instance, certain applications, such as human target localization defined in the 802.11bf standard, require a sensing interval of less than $20~\text{ms}$ \cite{TGbf_nist, Du_comst}.
We focus on a coexistence scenario that includes one 802.11ax AP and a growing number of 802.11bf APs. From Figure~\ref{TSUS_1_33}, we observe that reducing the sensing interval from $1000~\text{ms}$ to $10~\text{ms}$, significantly decreases the average 802.11bf latency. For example, in the presence of one 802.11bf AP and one 802.11ax AP, shortening the interval from $1000~\text{ms}$ to $100~\text{ms}$ reduces the latency from approximately $8~\text{ms}$ to $4~\text{ms}$. This improvement arises because shorter intervals enable more frequent sensing attempts, allowing the 802.11bf AP to secure channel access more regularly and thus lower the overall sensing latency.
However, as the number of 802.11bf APs increases, the average latency gradually rises, particularly at longer intervals. This indicates a compounded effect of additional sensing traffic and multiple contending APs, which collectively lead to increased waiting times and, consequently, higher latency. 

Notably, with a sensing interval of $10~\text{ms}$, the average latency remains almost constant as the number of 802.11bf APs increases. This is due to the exclusion of sensing failure events from the average latency calculation, where sensing attempts that cannot be completed before the next sensing interval are dropped.
Table~\ref{tab:bf_ax_failure} presents the sensing failure percentages across different sensing intervals. We define a sensing failure for an 802.11bf AP as an event where a sensing instance cannot be completed before the next sensing interval begins due to collisions or channel occupation by other APs. In such cases, the sensing attempt is dropped and excluded from the delay calculations. As shown in Table~\ref{tab:bf_ax_failure}, sensing failures are more prominent for the $10~\text{ms}$ sensing interval, with failure rates exceeding $50\%$ even when the number of 802.11bf APs is three or more.

\begin{table}[htbp]
    \centering
    \caption{Sensing Failure Across Different Sensing Intervals}
    \label{tab:bf_ax_failure}
    \resizebox{\columnwidth}{!}{%
    \begin{tabular}{|l|c|c|c|c|c|}
        \hline
         \textbf{Number of BSS}& \multicolumn{5}{|c|}{\textbf{Sensing Interval}} \\ \hline
                              & \textbf{10 ms} & \textbf{50 ms} & \textbf{100 ms} & \textbf{500 ms} & \textbf{1000 ms} \\ \hline
        1 bf \& 1 ax          & 34.91\%           & 0.27\%           & 0.03\%            & 0.00\%            & 0.00\%             \\ \hline
        2 bf \&  1 ax          & 48.83\%            & 1.38\%           & 0.68\%            & 0.01\%            & 0.03\%             \\ \hline
        3 bf \&  1 ax          & 60.74\%            & 2.42\%            & 0.91\%            & 0.09\%            & 0.02\%             \\ \hline
        4 bf \&  1 ax          & 68.94\%            & 3.44\%            & 1.57\%             & 0.14\%            & 0.04\%             \\ \hline
        5 bf \&  1 ax          & 77.51\%            & 4.10\%            & 1.58\%             & 0.20\%            & 0.07\%             \\ \hline
    \end{tabular}
    }
\end{table}

 Figure~\ref{TSUS_1_7} shows that the throughput of 802.11ax significantly declines as the sensing interval becomes shorter.
 For instance, when we have one 802.11bf and one 802.11ax network for a 10~ms sensing interval, the throughput of 802.11ax drops to 50 Mbps compared to 58 Mbps for 1000~ms. This reduction becomes more pronounced as the number of 802.11bf APs increases, with throughput dropping to 30 Mbps for 10~ms intervals in case of having five 802.11bf networks. This degradation occurs because more frequent sensing leads to increased contention for the channel, reducing the airtime available for 802.11ax transmissions.
 
 In summary, while the latency performance of 802.11bf sensing applications benefits from shorter sensing intervals, it comes at the cost of reduced 802.11ax throughput. This trade-off emphasizes the importance of balancing sensing interval durations to optimize coexistence performance in shared networks.

\begin{figure}[h!]
\centering\includegraphics[width=0.49\textwidth,trim = 5mm 5mm 5mm 0mm,clip]{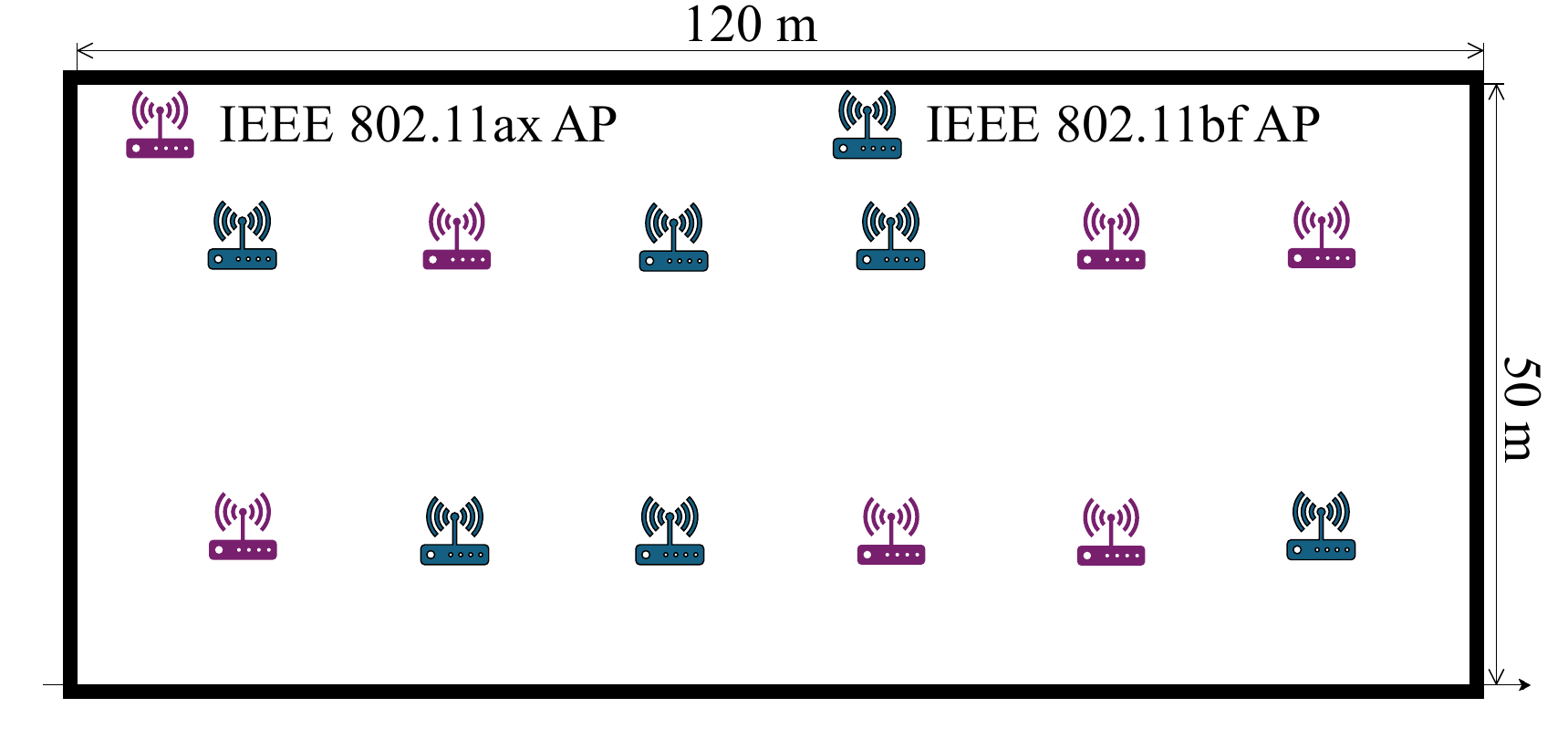}
	\caption{Layout of indoor office scenario.}
	\label{fig:layout_office}
\end{figure}

\subsection{Effect of number of antenna element}
\label{subsection:antenna-element}
\begin{figure*}[!ht]

    \centering
    \subfigure{
\includegraphics[width=0.312\textwidth,trim = 0mm 0mm 0mm 0mm,clip]{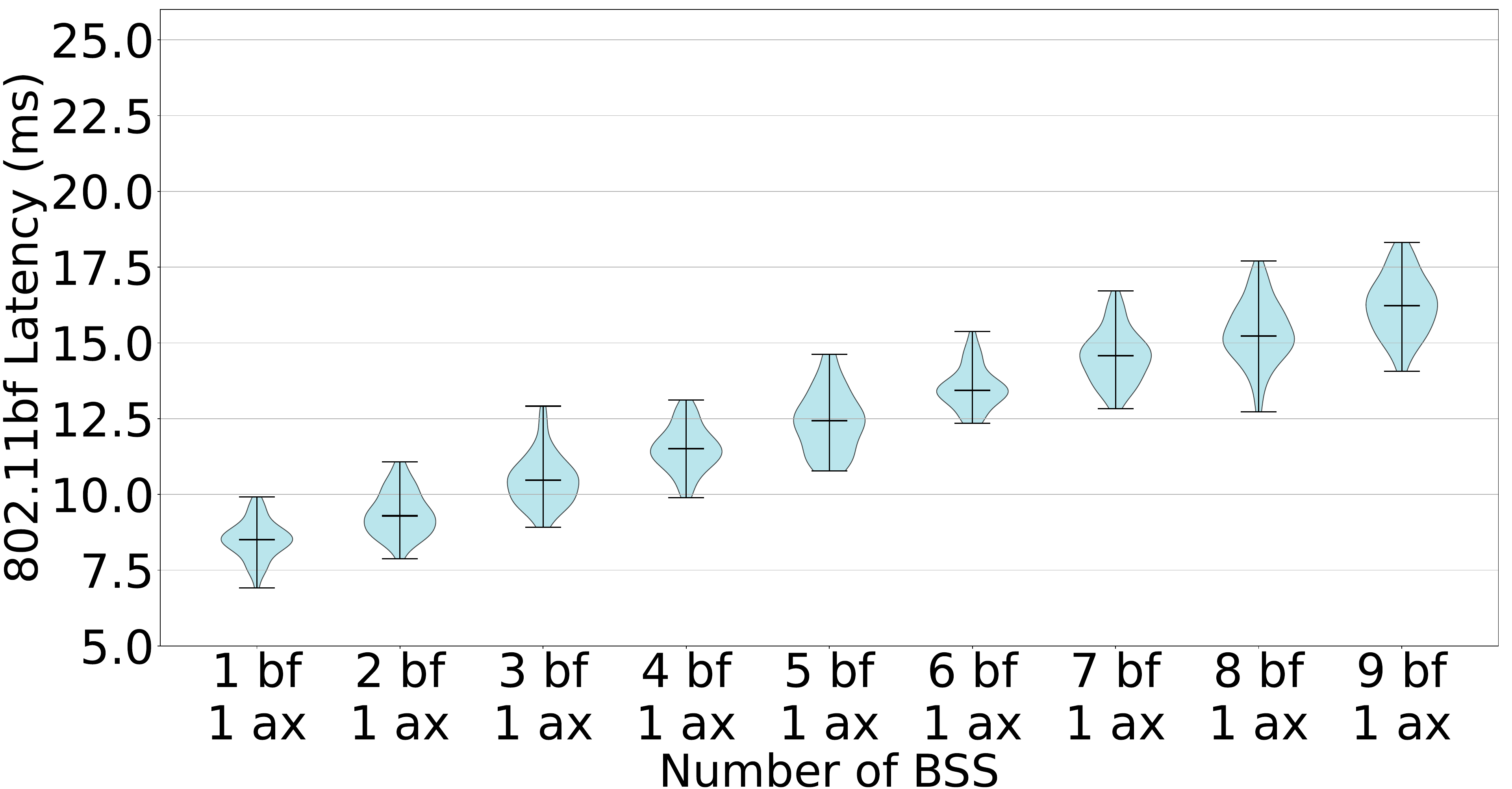}
    \label{4_l}
    }
    \addtocounter{subfigure}{-3}
    \subfigure{
\hspace{-1mm}\includegraphics[width=0.312\textwidth,trim =0mm 0mm 0mm 0mm,clip]{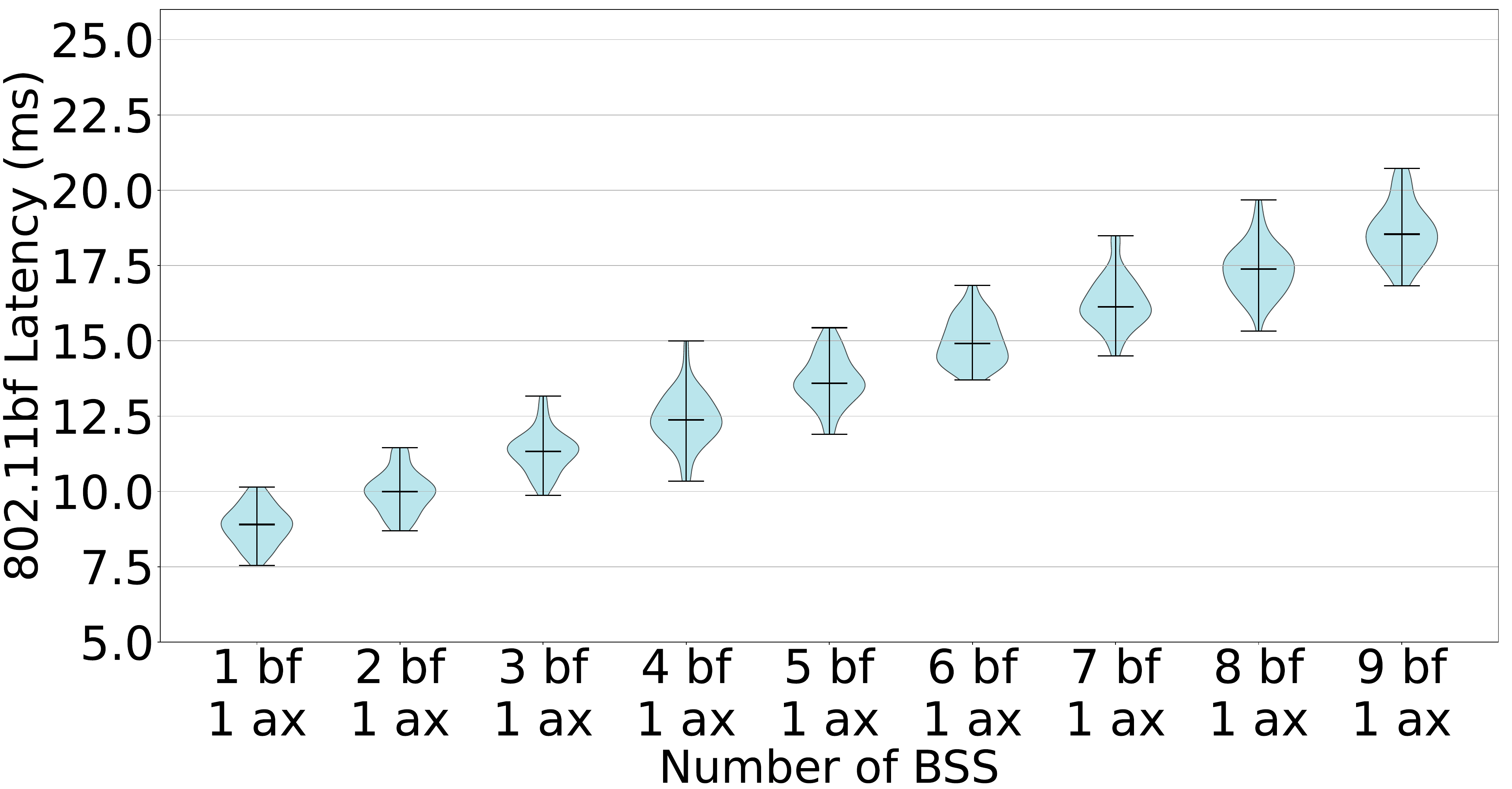}
    \label{8_l}
    }
    \subfigure{
    \hspace{-1mm}
    \includegraphics[width=0.312\textwidth,clip]{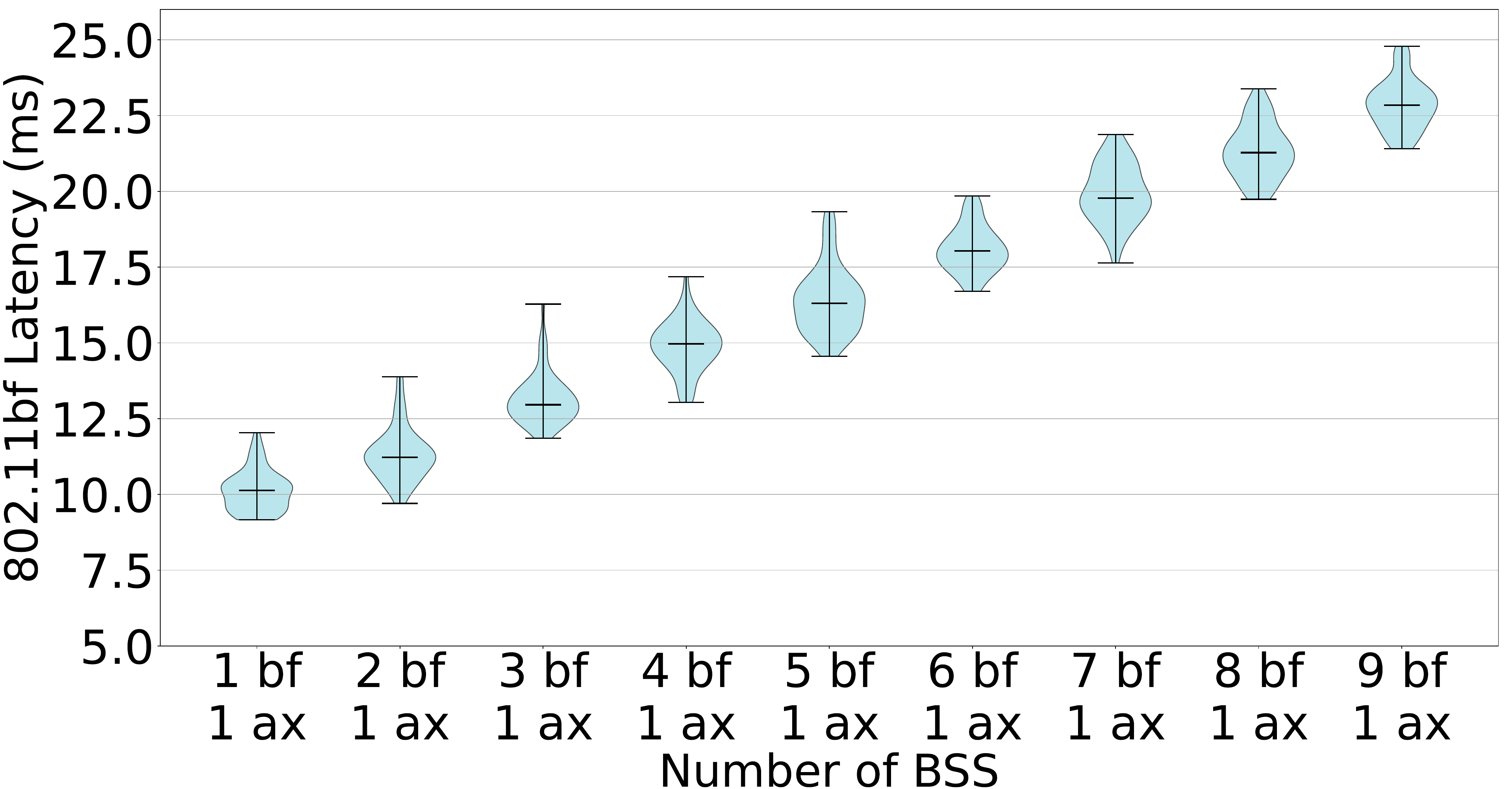}
    \label{16_l}
    }
\end{figure*}
\begin{figure*}[!ht]
    \centering
    \subfigure[{4 Antenna Elements}]{
    \hspace{0.5mm}
    \includegraphics[width=0.312\textwidth,trim = 0mm 0mm 0mm 0mm,clip]{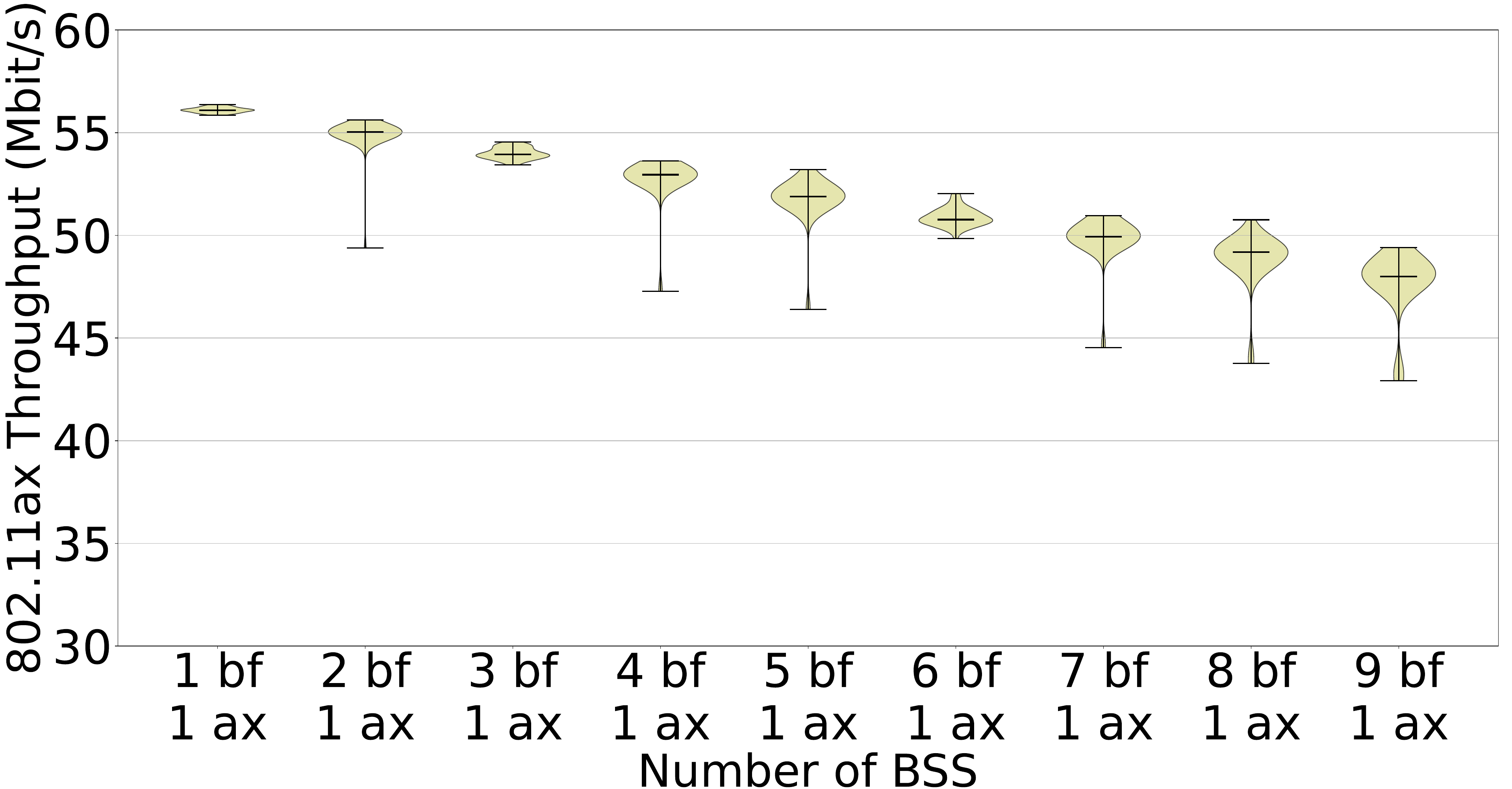}
    \label{4_th}
    }
    \subfigure[{8 Antenna Elements}]{
\hspace{-1mm}\includegraphics[width=0.312\textwidth,trim =0mm 0mm 0mm 0mm,clip]{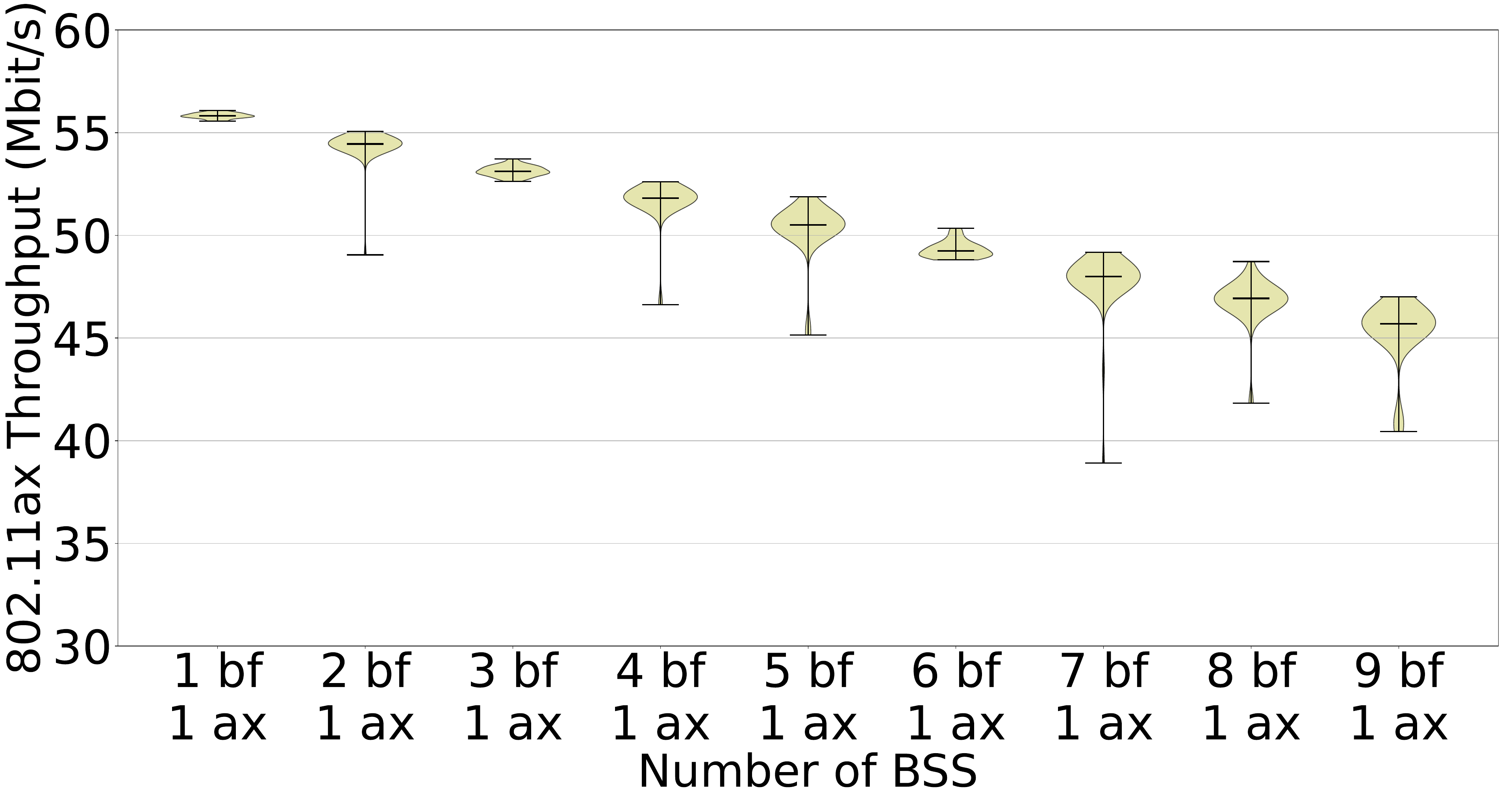}
    \label{8_th}
    }
    \subfigure[{16 Antenna Elements}]{
    \includegraphics[width=0.312\textwidth,clip]{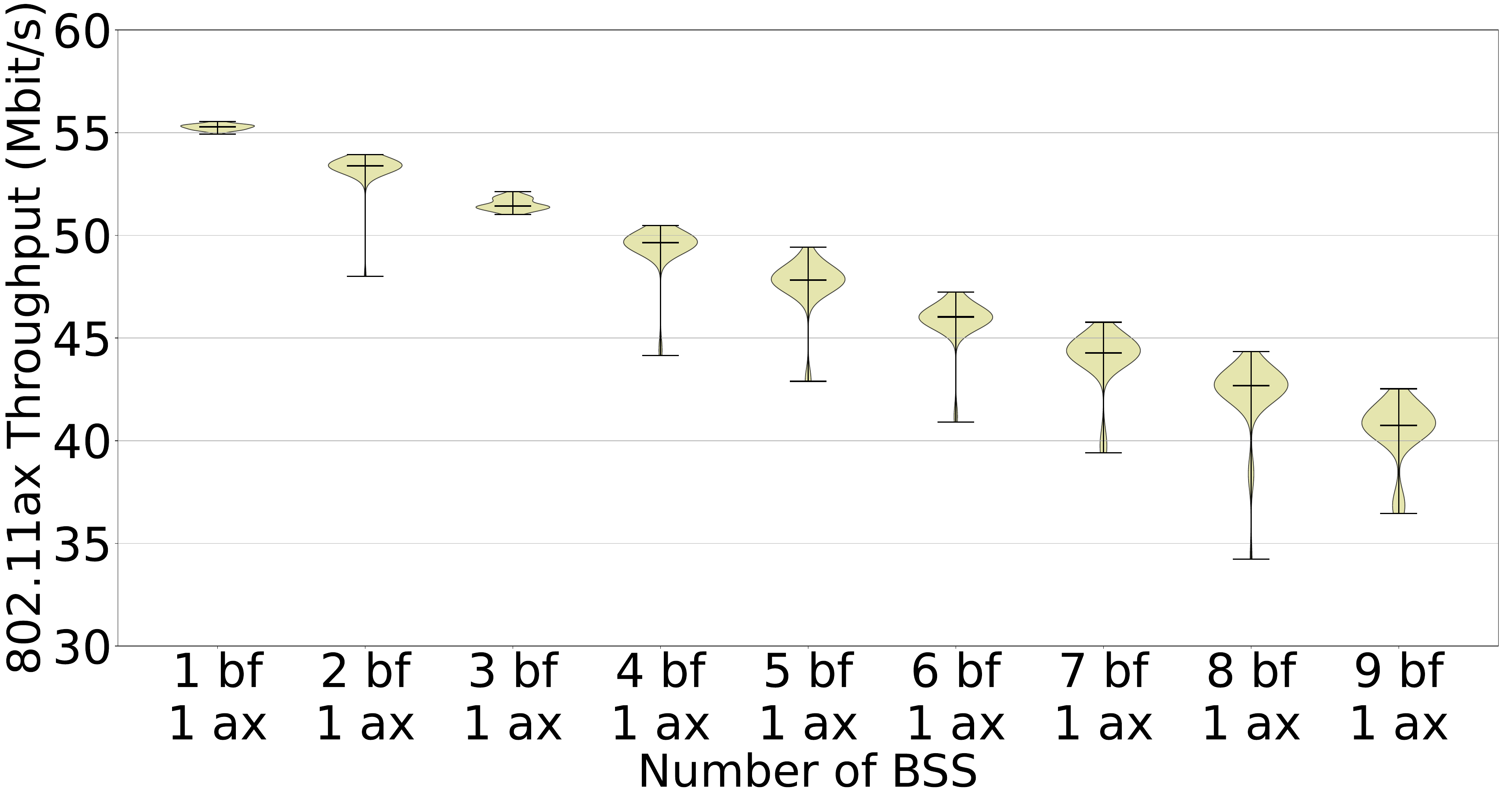}
    \label{16_th}
    }
    \caption{802.11bf Latency and 802.11ax Throughput Distribution for various number of antenna elements.}
\label{result_antenna}
\end{figure*}

Having more antenna elements in 802.11bf networks enables richer CSI acquisition (as shown in Equation~\eqref{csi_size}), potentially improving sensing accuracy and performance \cite{Du_comst}. However, this benefit comes at the expense of increased reporting overhead and, consequently, higher latency. In this subsection, we quantify this trade-off by examining how scaling up the number of antenna elements at both the AP and STA affects coexistence with 802.11ax networks.

Until now, we have assumed the use of a single omnidirectional antenna for both the AP and STA in 802.11bf networks. In this subsection, we examine the impact of increasing the number of antenna elements at both the AP and STA. Specifically, we configure the STA with a 2x2 antenna array and vary the number of antenna elements at the AP to  4x4, 8x8, and 16x16, while keeping the 802.11ax omnidirectional.

Figure~\ref{result_antenna} presents the results of this study. The first row illustrates 802.11bf latency, while the second row shows the corresponding 802.11ax throughput. In all subfigures, increasing the number of 802.11bf APs leads to a performance trend similar to that observed in Figure~\ref{fig:scenario-1}. However, comparing subfigures (a), (b), and (c) reveals that scaling the AP’s antenna configuration from 4x4 to 8x8 slightly increases 802.11bf latency and decreases 802.11ax throughput. This effect becomes even more pronounced at 16x16. The root cause lies in the enlarged CSI reporting size, as described in \eqref{csi_size}. With more antenna elements, the CSI feedback is more data-intensive, prolonging the reporting phase and subsequently affecting both 802.11bf latency and 802.11ax throughput under coexistence conditions.

\subsection{Performance Evaluation in Indoor Office Scenario}

In this Subsection, we consider the indoor office scenario defined by the 3GPP TR 38.901 standard \cite{3GPP2} serves as a realistic environment for evaluating the coexistence of 802.11bf sensing and 802.11ax communication networks. We consider typical office conditions—featuring corridors, open spaces, and partitioned areas—where line-of-sight (LoS) and non-line-of-sight (NLoS) propagation paths coexist. The layout, illustrated in Figure~\ref{fig:layout_office}, encompasses an area of $120 \times 50$ meters, within which 12 APs are uniformly distributed at an inter-AP spacing of $20$ meters. To systematically assess the interplay between sensing and communication, we vary the ratio of 802.11bf (sensing) APs to  802.11ax (communication) APs, thereby enabling us to explore different coexistence conditions under varying densities of each technology. 
Additionally, we investigate the impact of larger channel bandwidths (e.g., 80~MHz or 160~MHz) on practical indoor sensing scenarios. Employing wider bandwidths improves the spatial and temporal resolution of the acquired CSI, thereby enabling more precise sensing tasks such as human target localization or object detection \cite{TGbf_nist}.

\begin{figure*}[!ht]
    \centering
    \subfigure[{20~MHz}]{
    \hspace{0.5mm}
    \includegraphics[width=0.314\textwidth,trim = 0mm 0mm 0mm 0mm,clip]{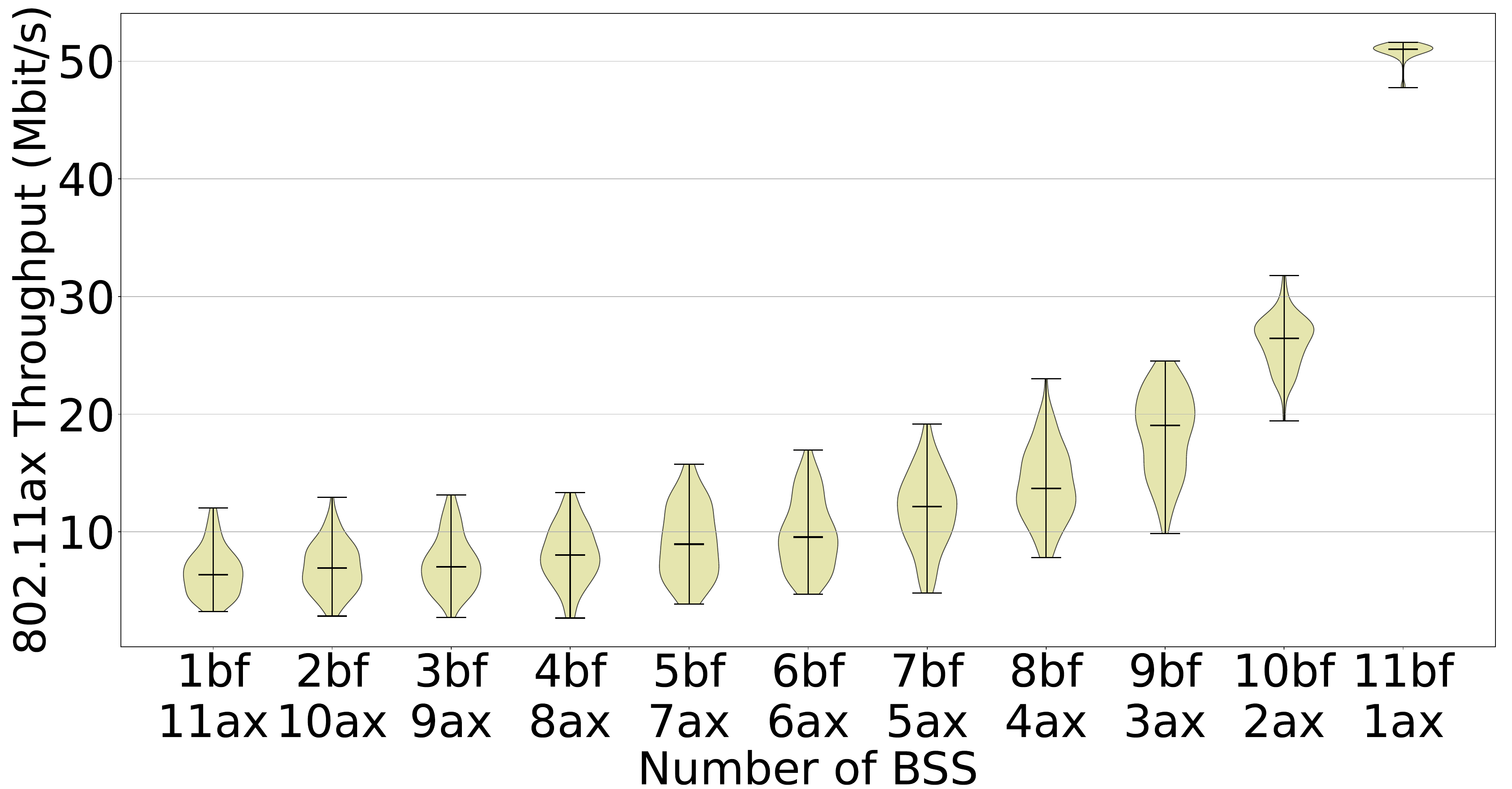}
    \label{20_th}
    }
    \subfigure[{80~MHz}]{
\hspace{-1mm}\includegraphics[width=0.314\textwidth,trim =0mm 0mm 0mm 0mm,clip]{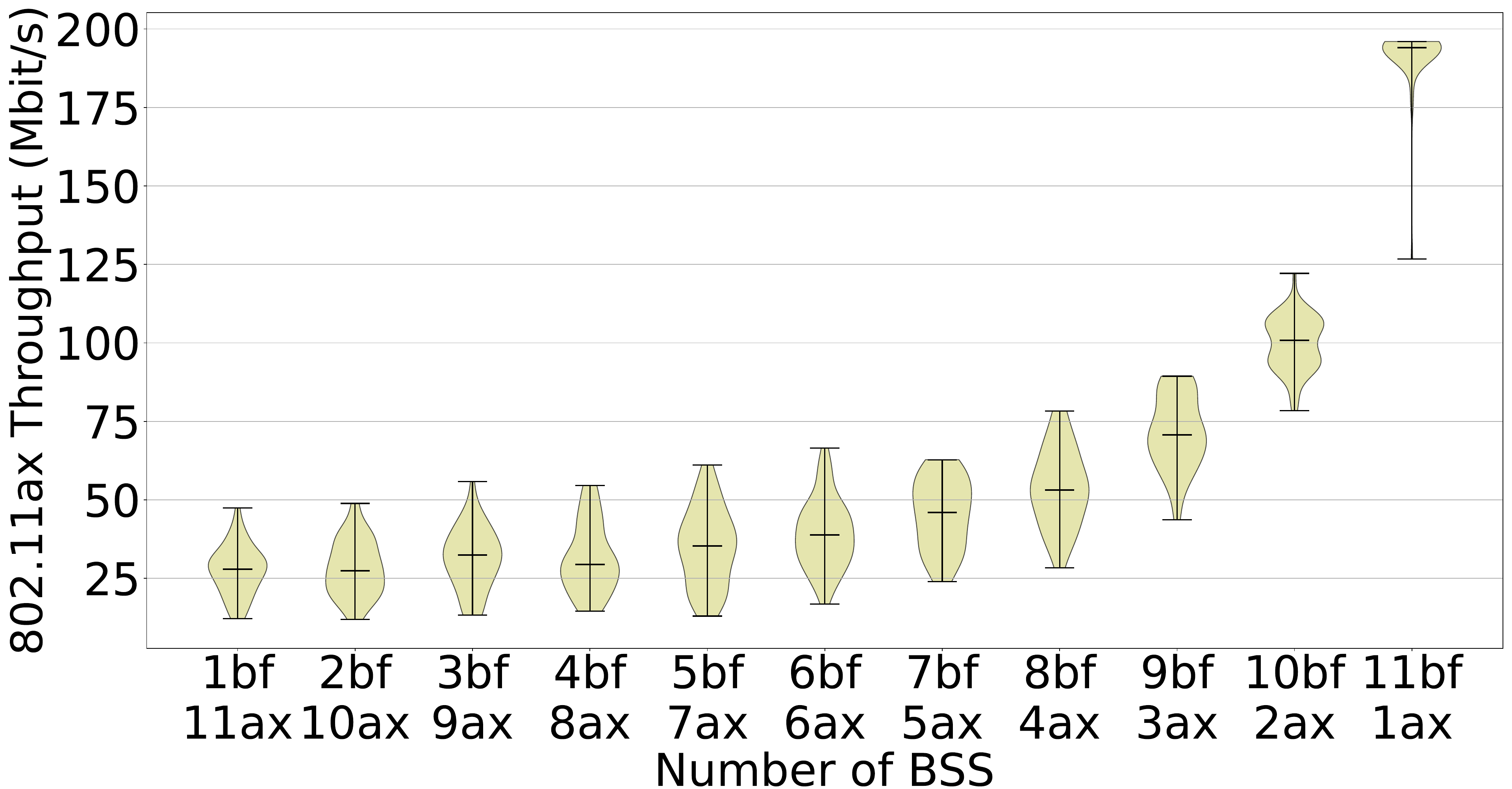}
    \label{80_th}
    }
    \subfigure[{160~MHz}]{
    \includegraphics[width=0.314\textwidth,clip]{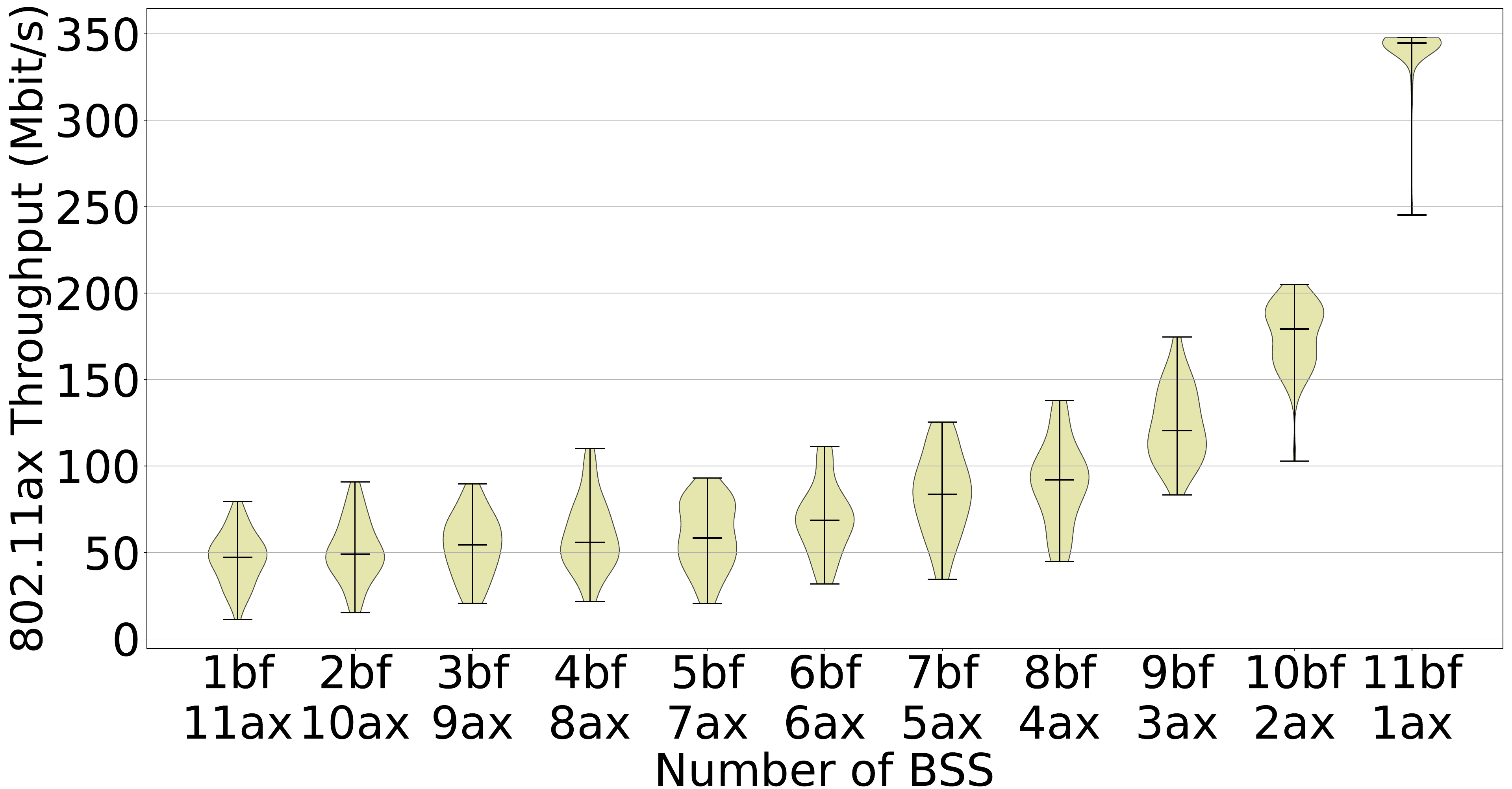}
    \label{160_th}
    }
    \caption{802.11bf Latency and 802.11ax throughput distribution for various channel bandwidth.}
\label{result_office}
\end{figure*}

Figure~\ref{result_office} shows the latency and throughput distributions for 802.11bf and 802.11ax coexistence under 20~MHz, 80~MHz, and 160~MHz channel bandwidths. The violin curves reveal that as the ratio of 802.11bf APs increases relative to 802.11ax APs, the overall latency distribution shifts toward lower values. For instance, at 20~MHz, the average latency for 802.11bf drops from about 33~ms when only one 802.11bf AP coexists with eleven 802.11ax APs, to approximately 8~ms when eleven 802.11bf APs coexist with one 802.11ax AP. This trend is even more pronounced at higher bandwidths, with the median latency decreasing from around 12~ms to 7~ms at 80~MHz, and from about 8~ms to 5.5~ms at 160~MHz.

This improvement stems from the fact that 802.11bf APs engage in shorter, more frequent sensing operations—polling, sounding, and reporting—especially when richer CSI, enabled by wider bandwidths, provides more accurate and efficient sensing with less overhead per operation. Wider bandwidths also enhance the resolution and quality of CSI, enabling more precise sensing and shorter reporting periods. Conversely, 802.11ax APs, operating with full-buffered traffic, hold the channel longer during each TXOP, potentially up to 5,484~ms, thus increasing latency for 802.11bf when they dominate the network. By visualizing results as violin curves, we clearly see how the latency distribution narrows and shifts downward as the ratio of sensing APs grows,

For throughput analysis, we observe that under scenarios where 802.11ax APs are higher relative to 802.11bf APs, channel time becomes heavily contested, resulting in lower per-network throughput. As the ratio shifts towards more  802.11bf APs, contention among 802.11ax APs decreases, allowing each network to utilize available airtime more effectively and thereby enhancing throughput. This improvement is further amplified by wider channel bandwidths, which provide greater capacity and reduce the number of competing 802.11ax APs. For instance, at a 20~MHz bandwidth, 802.11ax throughput increases from approximately 5Mbps with a 1:11 ratio of 802.11bf to 802.11ax APs, up to 50Mbps at an 11:1 ratio. Similarly, at 80~MHz and 160~MHz bandwidths, throughput rises significantly as the ratio of 802.11bf APs increases, reaching up to 200~Mbps and nearly 350~Mbps, respectively. 

 For throughput, the results show that for 20~MHz channel bandwidth, the mean throughput of 802.11ax starts at approximately 5~Mbps when the ratio of 802.11bf APs to 802.11ax APs is low (1:11) and increases steadily to 50~Mbps as the ratio increases to 11:1. For 80~MHz, the mean throughput begins higher, starting at around 25~Mbps and increasing to 200~Mbps. Similarly, for 160~MHz, the throughput exhibits a significant rise, beginning at approximately 50~Mbps and reaching almost 350~Mbps. This is because reducing the number of 802.11ax APs sharing the channel leads to an increase in the mean throughput for each network. In particular, the available channel time is divided among fewer APs with saturated traffic, leaving more time for individual networks to transmit.

\begin{figure*}[!ht]

    \centering
    \subfigure{
    \hspace{0.5mm}
    \includegraphics[width=0.312\textwidth,trim = 0mm 0mm 0mm 0mm,clip]{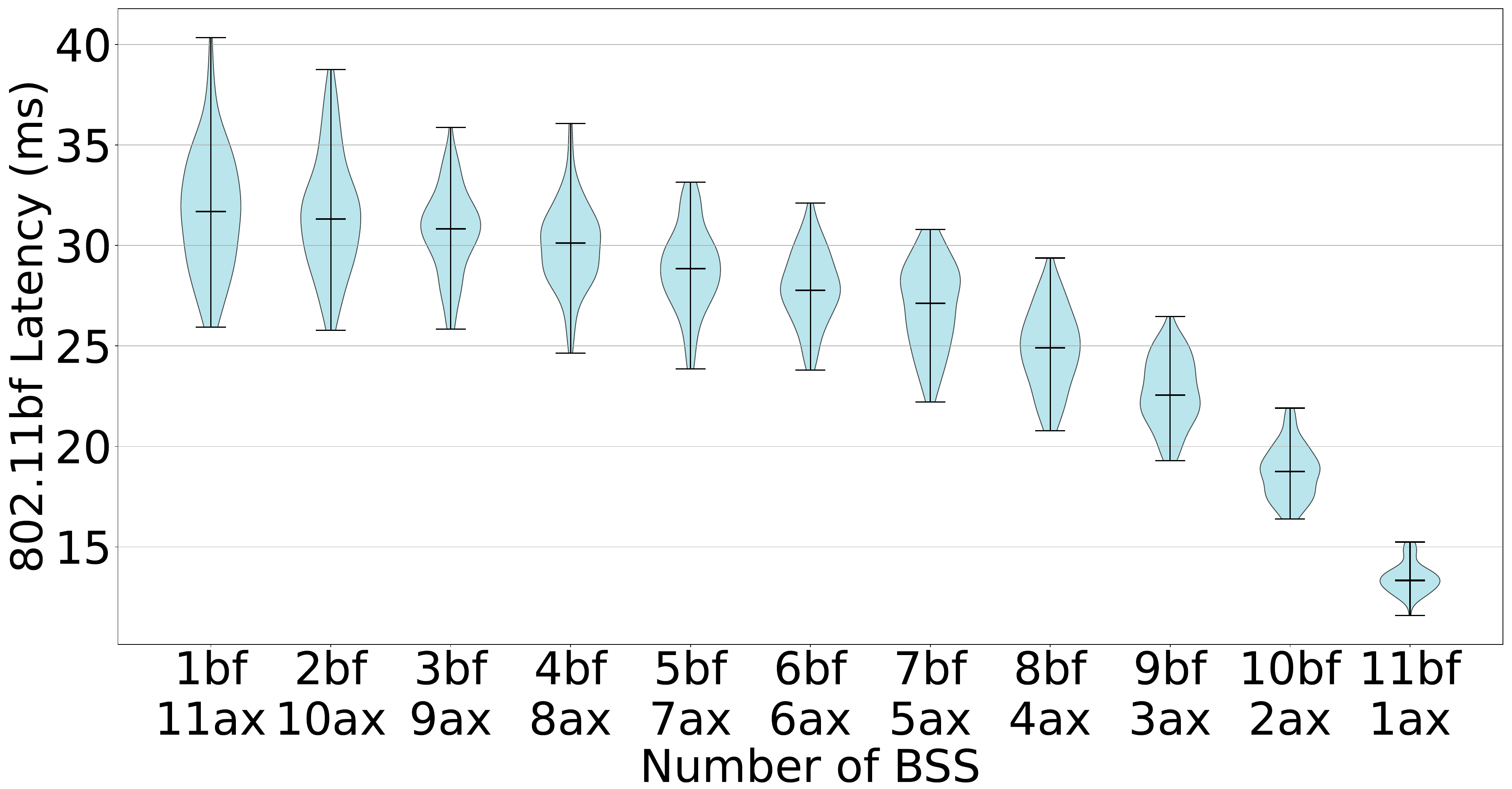}
    \label{20_l}
    }
    \addtocounter{subfigure}{-3}
    \subfigure{
\hspace{-1mm}\includegraphics[width=0.312\textwidth,trim =0mm 0mm 0mm 0mm,clip]{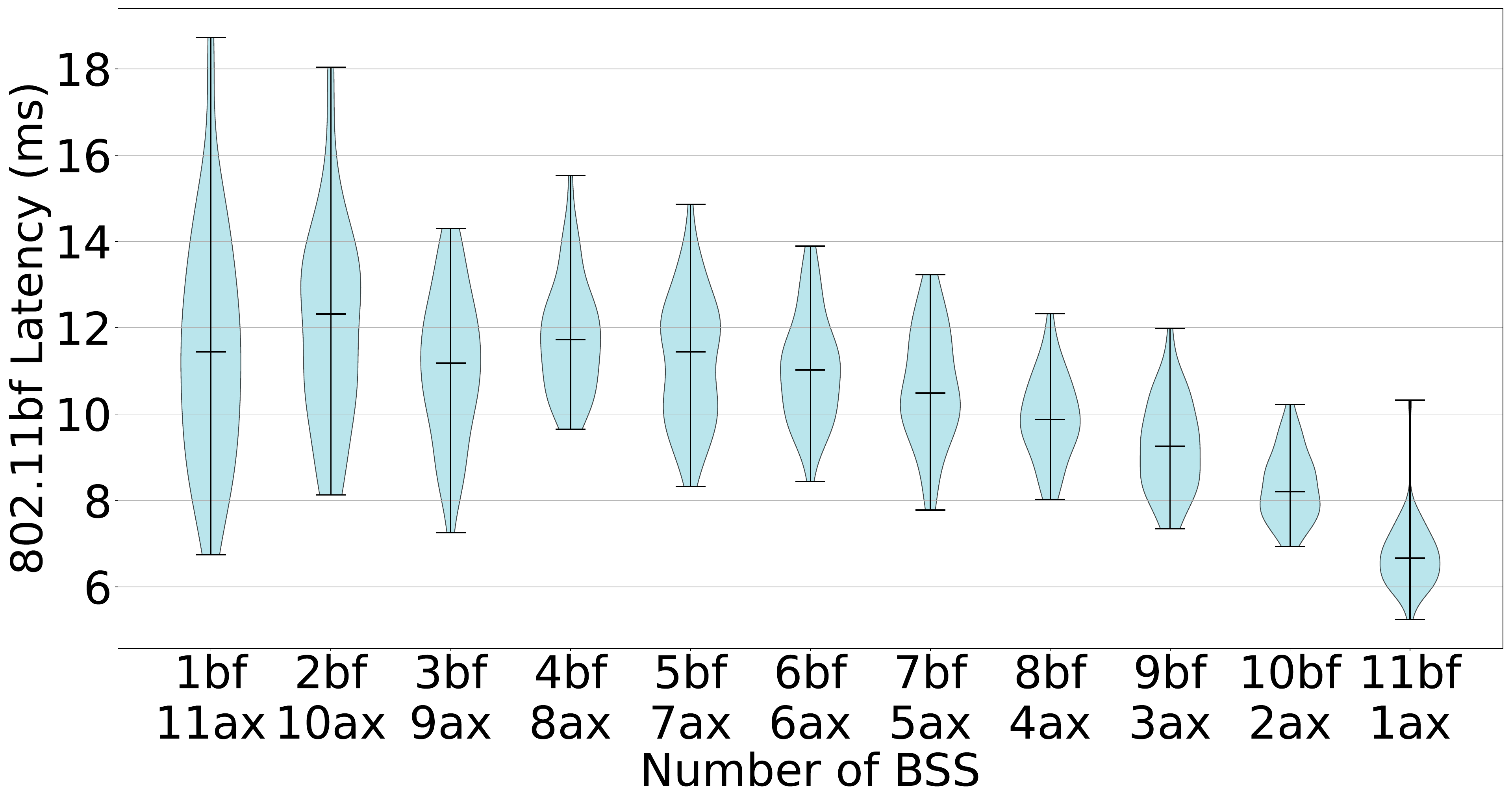}
    \label{80_l}
    }
    \subfigure{
    \includegraphics[width=0.312\textwidth,clip]{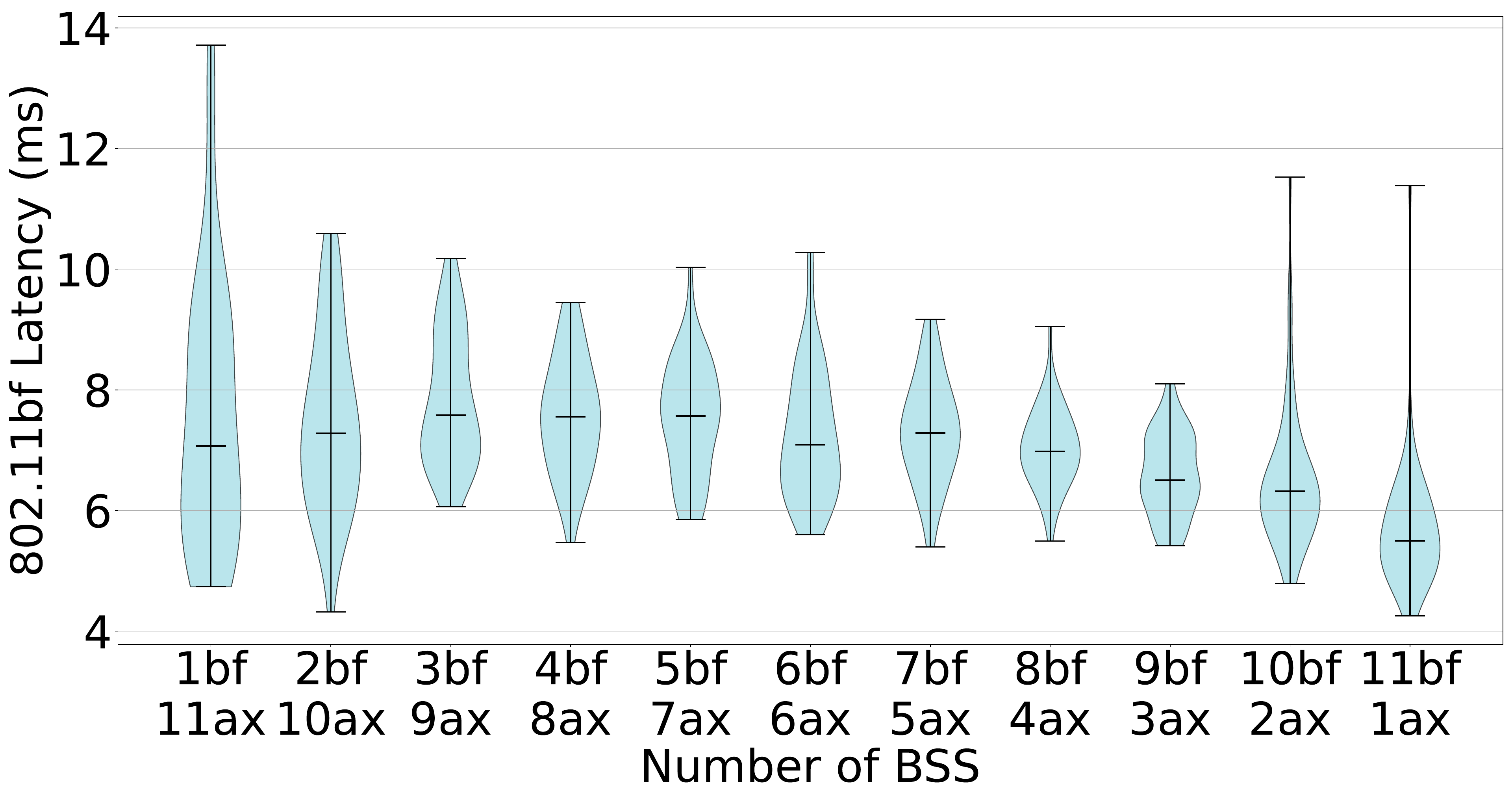}
    \label{160_l}
    }
\end{figure*}
\section{Conclusions}\label{sec:Conclusions}
In this paper, we provide key insights into how 802.11bf sensing networks coexist with legacy 802.11ax. By developing an analytical framework and conducting detailed simulations in \mbox{ns-3}, we demonstrated that while 802.11bf can operate effectively alongside 802.11ax, channel contention significantly affects performance, leading to increased sensing delays and reduced communication throughput. Our analysis quantifies these trade-offs and illustrates how factors such as node density, sensing intervals, and MAC configurations influence the coexistence dynamics. Moreover, by examining various choices for sensing intervals, channel bandwidth, and antenna configurations, and validating in a 3GPP TR 38.901 office channel scenario, we confirmed the practical impact of these design considerations on real-world deployments.
In future work, we plan to incorporate the intelligent channel access mechanism into 802.11bf networks to further mitigate sensing delays, reduce protocol overhead, and minimize adverse effects on other Wi-Fi networks.
\bibliographystyle{IEEEtran}
\bibliography{icc_radcom}
\appendix
\label{appedix}
\section*{Expressions for $d_I$, $d_S$, and $d_C$}

As specified in Equation \eqref{overal_delay}, to compute the overall channel access delay of a tagged 802.11bf AP, we need to calculate $\mathcal{F}$, which is the average duration that the AP stays in the backoff state before decrementing its counter. The time elapsed in each backoff state depends on the channel state of the current new time slot as well as the previous time slot. We define $d_I$, $d_S$, and $d_C$ as the duration that the AP stays in the backoff state when the channel status at the beginning of that slot is idle, involved in a successful event, or experiencing a collision, respectively.   

In case the channel is idle, the tagged AP remain in this state for one time slot duration, i.e. $d_I=1$. For successful sensing operation of 802.11bf sensing, the delay $d_S$ can be expressed by using the geometric distribution as
\begin{equation}
\begin{aligned}
    d_S=&{d_I}+\frac{P_{t,ax}P_{s,ax}(1-P_{t,bf})T_{s,ax}}{1-P_{ss}}\\&+\frac{P_{t,bf}P_{s,bf}(1-P_{t,ax})T_{s,bf}}{1-P_{ss}}{}.
    \end{aligned}
\end{equation}
Where $P_{ss}$ denotes the probability of successive successful sensing events, i.e., $P_{ss}=1/W_0$, and $W_0$ is the minimum contention window size at backoff stage $0$. The expressions of $P_{t,bf}$, $P_{t,ax}$, $P_{s,bf}$, and $P_{s,ax}$ are given in equations \eqref{coll1_prob} and \eqref{succ_prob}, respectively. Further, the expressions of $T_{s,ax}$, $T_{s,bf}$ are provided already in equation \eqref{eq_24}.

Further, the delay experienced when the AP enter the backoff state and encounter a collision, $d_C$, is given by
\begin{equation}\label{dc_appendix}
d_C=\left(\sum_{i=0}^L i P_{c c}^i\right) \Gamma+\frac{P_{c s}}{1-P_{c c}} d_S+\frac{P_{c i}}{1-P_{c c}} d_I.
\end{equation}
Where $\sum_{i=0}^L i P_{c c}^i$ represents the average number of consecutive collisions. $P_{cs}$ is the probability of transitioning from a collision to a successful sensing event, and $P_{ci}$ is the probability of transitioning from a collision to an idle state. Finally, $\Gamma$ represents the total time elapsed in both inter and intra-technology collisions and can be expressed as:
\begin{equation}
\begin{aligned}
 \Gamma &= P_{t,ax}(1-P_{s,ax})(1-P_{t,bf})T_{c,ax}\\&+P_{t,bf}(1-P_{s,bf})(1-P_{t,ax})T_{c,bf}+\overline{T_{c}}(P_{t,ax}P_{s,ax}\\&P_{t,bf}P_{s,bf}+P_{t,ax}P_{s,ax}P_{t,bf}(1-P_{s,bf})+P_{t,ax}(1-P_{s,ax})\\&P_{t,bf}P_{s,bf}+P_{t,ax}(1-P_{s,ax})P_{t,bf}(1-P_{s,bf})).  
\end{aligned}  \nonumber  
\end{equation}
The expressions of $T_{c,ax}$ and $T_{c,bf}$ are given in the equation \eqref{eq_242}.

To calculate the probability of $P_{cs}$ and $P_{cs}$, we need to define the joint probability mass function (PMF) of a number of colliding APs. Let $Q(n_{ax}, n_{bf})$ be the probability that $n_{ax}$ 802.11ax APs and $n_{bf}$ 802.11bf APs collide in the current time slot, which can be expressed as:
\begin{equation}\label{q_pmf}
\begin{aligned}
Q(n_{ax}, n_{bf}) =& \binom{N_{ax}}{n_{ax}} \tau_w^{n_{ax}} (1 - \tau_{ax})^{N_{ax} - n_{ax}}        \\&\times\binom{N_{bf}}{n_{bf}} \tau_{bf}^{n_{bf}} (1 - \tau_{bf})^{N_{bf} - n_{bf}}.
\end{aligned}
\end{equation}
Now, utilizing Equation \eqref{q_pmf}, we can express the $P_{ci}$ as following
\begin{equation}\label{pci_appendix}
\begin{aligned}
P_{ci} = \sum_{n_{bf}=2}^{N_{bf}-1} \sum_{n_{ax}=0}^{N_{ax}} & \bigg[ Q(n_{ax}, n_{bf}) \left( 1 - \frac{1}{\overline{CW_{ax}}} \right)^{n_{ax} }\\&\left( 1 - \frac{1}{\overline{CW_{bf}}} \right)^{n_{bf} }\bigg],
\end{aligned}
\end{equation}
where $\overline{CW_{bf}}$ and $\overline{CW_{ax}}$ denote the average backoff window size for all backoff stages defined in 802.11bf and ax standards, respectively. Similarly, the expression of $P_{cs}$ is 

\begin{equation}\label{pcs_appendix}
\begin{aligned}
P_{cs}=&\sum_{n_{bf}=2}^{N_{bf}-1}\sum_{n_{ax}=0}^{N_{ax}}  Q(n_{ax}, n_{bf}) n_{bf} \left(\frac{1}{\overline{C W_{bf}}}\right)\\&\times\left(1-\frac{1}{\overline{C W_{bf}}}\right)^{n_{bf}-1}\left(1-\frac{1}{\overline{C W_{ax}}}\right)^{n_{ax}}+\\&\sum_{n_{ax}=1}^{N_{ax}}  Q(1, n_{ax}) \left(\frac{1}{\overline{C W_{bf}}}\right)\times\left(1-\frac{1}{\overline{C W_{ax}}}\right)^{n_{ax}}.
\end{aligned}
\end{equation}
Finally, the probability of transitioning from a collision to a collision state is:
\begin{equation}\label{pcc_appendix}
P_{cc}=1-P_{ci}-P_{cs}.
\end{equation}
Substituting the equations \eqref{pci_appendix}, \eqref{pcs_appendix}, and \eqref{pcc_appendix} in Equation \eqref{dc_appendix}, we can obtain the final expression of $d_C$.

\end{document}